\documentclass[10pt,letterpaper]{article}

\usepackage[
    top=0.80in,
    bottom=0.80in,
    left=0.75in,
    right=0.75in
]{geometry}

\usepackage{newtxtext}
\usepackage{newtxmath}

\usepackage{setspace}
\usepackage{microtype} 
\usepackage{etoolbox} 

\usepackage{amsmath}
\usepackage{graphicx}
\usepackage{subcaption}
\usepackage{booktabs}
\usepackage{multirow}
\usepackage{siunitx}

\usepackage{array}
\usepackage{tabularx}

\usepackage{xurl}
\usepackage{xspace}
\newcommand{\etal}{\textit{et al.}\xspace}

\usepackage[numbers,sort&compress]{natbib}

\usepackage{titlesec}

\titleformat{\section}
  {\normalfont\fontsize{10}{12}\selectfont\bfseries}
  {\thesection.}
  {0.4em}
  {}

\titleformat{\subsection}
  {\normalfont\fontsize{10}{12}\selectfont\bfseries}
  {\thesubsection.}
  {0.4em}
  {}

\titleformat{\subsubsection}
  {\normalfont\fontsize{10}{12}\selectfont\bfseries}
  {\thesubsubsection.}
  {0.4em}
  {}

\titlespacing*{\section}{0pt}{12pt}{6pt}
\titlespacing*{\subsection}{0pt}{10pt}{4pt}
\titlespacing*{\subsubsection}{0pt}{8pt}{4pt}

\usepackage{float}

\AtBeginEnvironment{tabular}{\singlespacing}
\begin{document}

\doublespacing


\begin{center}

{\fontsize{12}{18}\selectfont\bfseries
Physics-Informed Latent Neural ODEs for Reduced-Order Modeling of Hydrogen, Ammonia, and Methane Combustion Kinetics
\par}

\vspace{10pt}

{\fontsize{10}{12}\selectfont
Jay Patel, Shubhangi Bansude$^{*}$
\par}

\vspace{6pt}

{\fontsize{10}{12}\selectfont\itshape
Department of Mechanical Engineering,
Indian Institute of Technology Gandhinagar,
Gandhinagar, Gujarat, India
\par}

\end{center}

\vspace{8pt}

\noindent{\bfseries Abstract}

Detailed chemical kinetic mechanisms accurately describe combustion chemistry
but are computationally expensive in reacting-flow simulations because of their
high dimensionality and numerical stiffness. This study develops an end-to-end physics-informed autoencoder neural ordinary differential equation (PI-AE-NODE) framework to accelerate integration of combustion kinetics. An encoder compresses the thermochemical state into a low-dimensional latent space, a Neural ODE advances this latent state continuously in time, and a decoder reconstructs the temperature and all
species mass fractions. Elemental conservation, total mass conservation, and
enthalpy consistency are embedded in the training objective, so that the
learned latent dynamics remain physically consistent rather than purely
data-fitted. The framework is applied to zero-dimensional H$_2$/air,
NH$_3$/air, and CH$_4$/air autoignition using mechanisms with 9, 24, and 52
species (21, 100, and 323 reactions), whose 10-, 25-, and 53-dimensional
states are represented by 5, 8, and 10 latent variables. Relative to detailed
kinetics, the mean ignition delay errors are 2.20\%, 1.75\%, and 3.28\%,
respectively. For H$_2$/air, the physical constraints reduce this error from
6.19\% for an otherwise identical data-driven model and substantially lower
elemental and mass conservation errors near ignition. The models reproduce the
induction period, heat release, major and intermediate species evolution, and
post-ignition state at both training and unseen conditions. The learned latent dynamics permit efficient explicit integration and yield speedups of
up to 5.69$\times$, 3.96$\times$, and 26.00$\times$ over implicit integration
of the detailed mechanisms, with the largest gain for the largest mechanism.
PI-AE-NODE thus provides an accurate, physically consistent, and efficient
surrogate for detailed combustion chemistry across fuels of increasing
complexity.

\vspace{6pt}


\clearpage

\section{Introduction}
\label{sec:introduction}

Combustion is a fundamental energy conversion process that supports power
generation, transportation, industrial heating, chemical manufacturing, and
propulsion. The development of cleaner and more efficient combustion
technologies requires computational tools capable of predicting reacting-flow
behavior over broad operating conditions. Hydrogen and ammonia have received
considerable attention as carbon-free energy carriers, although their
combustion characteristics differ substantially from those of conventional
hydrocarbon fuels. Hydrogen exhibits high reactivity, high laminar flame
speeds, and wide flammability limits. In contrast, ammonia has a high ignition
temperature, low burning velocity, and complex nitrogen chemistry that governs
the formation of pollutants such as NO and N$_2$O \cite{Sun2025AmmoniaReview}.
Methane remains an important transitional fuel and is widely used as a
reference fuel for hydrocarbon combustion. Computational methods that
accurately and efficiently represent the distinct oxidation behavior of these
three fuels are therefore important for the design and optimization of
practical combustion systems.

Reacting-flow simulations with detailed chemistry remain computationally
demanding because combustion involves strongly coupled chemical reactions,
fluid motion, heat transfer, and mass transport over widely separated spatial
and temporal scales. Among these processes, the integration of detailed
chemical kinetics represents one of the principal computational bottlenecks.
Detailed reaction mechanisms may contain tens to hundreds of species and
hundreds to thousands of elementary reactions, resulting in large systems of
coupled nonlinear ordinary differential equations. These equations are
inherently stiff because the characteristic timescales of the underlying
chemical processes may differ by several orders of magnitude
\cite{Lu2009RealisticFuelChemistry,Bansude2026PhysicsGuidedML}. Resolving both
fast radical reactions and comparatively slow fuel conversion pathways
generally requires computationally expensive implicit integration methods. In
multidimensional simulations, the kinetic equations must be integrated
repeatedly in every computational cell and at every time step. Direct
integration of detailed chemistry can therefore account for a substantial
fraction of the total computational cost and may become prohibitive in
simulations involving turbulence, complex geometries, and broad operating
ranges \cite{Lu2009RealisticFuelChemistry}.

Several methods have been developed to reduce the computational cost of
detailed chemistry. Mechanism reduction techniques, including directed relation
graphs, quasi-steady-state approximations, computational singular
perturbation, and intrinsic low-dimensional manifolds, remove species and
reactions with limited influence or identify the dominant chemical timescales
\cite{Lu2005DRG,Lu2006QSSA,Valorani2006CSP,Maas1992ILDM}.
Chemistry tabulation methods, such as \textit{in situ} adaptive tabulation,
reduce repeated calculations by storing previously evaluated thermochemical
mappings and retrieving them when similar states are encountered
\cite{Pope1997ISAT}. Although these approaches can provide substantial
computational savings, they may require mechanism-specific development,
introduce approximation errors, retain residual stiffness, or impose
considerable memory requirements \cite{Chen2000ANNISAT}. Their accuracy may
also deteriorate outside the thermochemical domain for which the reduced or
tabulated representation was constructed. Hardware acceleration provides
another approach for reducing the cost of detailed chemistry. Cai
\etal\cite{Cai2026GPUSolver} developed a stiff ODE solver for combustion CFD
that uses analytical chemical kinetics and heterogeneous CPU and GPU
integration, demonstrating that detailed chemistry calculations can be
accelerated without removing species or reactions from the original mechanism.
However, the computational cost remains inherently constrained by the mechanism
size and numerical stiffness. For increasingly large and complex kinetic
systems, machine-learning surrogates provide an alternative by evaluating
learned source terms or thermochemical trajectories at substantially lower
computational cost.

Machine learning has consequently emerged as a promising approach for
accelerating detailed combustion chemistry \cite{Ihme2022CombustionML}. Early
neural network models approximated nonlinear mappings between thermochemical
states, chemical source terms, and compositions at later times. Ranade
\etal\cite{Ranade2019HybridChemistry} developed a hybrid chemistry framework
based on artificial neural networks for complex fuels. Chi
\etal\cite{Chi2021OnTheFlyANN} introduced a neural network framework that
constructs chemical reaction tabulations on the fly during direct numerical
simulations of premixed combustion. Zhang
\etal\cite{Zhang2022MultiscaleSampling} proposed a multiscale sampling and
preprocessing strategy to improve the accuracy and robustness of deep neural
networks for combustion kinetics. Their model was evaluated in configurations
ranging from homogeneous-reactor autoignition to multidimensional turbulent
flames, highlighting the importance of adequately sampling both major species
and reactive intermediates with low concentrations. Related approaches have focused on learning the temporal evolution or chemical source terms of reacting systems. Kumar and Echekki \cite{Kumar2024DeepONets}
developed a DeepONet framework that used a latent representation to predict
hydrogen and \textit{n}-dodecane oxidation over flexible time increments. Bai
\etal\cite{Bai2024MetaLearning} proposed a chemical decoupling method based on
a generalizable neural network trained using more than two million
thermochemical states and their corresponding source terms. These studies
demonstrated the potential of learned chemistry models for reducing the cost
of reacting-flow simulations. However, models that predict state changes or
source terms over discrete time increments may remain sensitive to the
distribution of the training data. Small instantaneous errors can also
accumulate when predictions are applied recursively or integrated over long
time horizons, resulting in inaccurate thermochemical trajectories.

Neural ordinary differential equations (Neural ODEs) provide an alternative
continuous-time formulation for learning chemical dynamics. In a Neural ODE, a
neural network represents the temporal derivative of the state, while a
differentiable numerical ODE solver advances the solution
\cite{Chen2018NeuralODE}. Neural ODEs can be trained using complete
trajectories, so that the interaction between the learned dynamics and
numerical integration is considered during optimization. The continuous-time
formulation also allows the chemical state to be advanced over time intervals
selected by the reacting-flow solver. Within an operator-splitting framework, a
Neural ODE surrogate can therefore replace the computationally expensive stiff
chemistry integration step while the flow and transport solvers remain
unchanged, which facilitates its incorporation into existing CFD algorithms
\cite{Bansude2023PMSR,Bansude2026PhysicsGuidedML}. Owoyele and
Pal \cite{Owoyele2022ChemNODE} introduced ChemNODE, in which separate neural
networks were trained for different species to integrate chemical kinetics
efficiently. Bansude \etal\cite{Bansude2023DataDrivenNODE} subsequently
developed a systematic strategy for data generation, transformation, and
training that used a single network for all species, and demonstrated accurate
prediction of H$_2$/air autoignition over a range of temperatures and
equivalence ratios. The performance of chemical kinetics Neural ODEs in
reacting and mixing systems was further investigated using a pairwise mixing
stirred reactor \cite{Bansude2023PMSR}. Saito \etal\cite{Saito2023AmmoniaNODE}
demonstrated the approach for NH$_3$/air combustion, while Dikeman
\etal\cite{Dikeman2022StiffnessReducedNODE} investigated Neural ODE models with
reduced stiffness for combustion kinetics. Although these surrogates have
demonstrated promising performance in zero-dimensional reactor configurations,
their integration into multidimensional CFD solvers is an active area of research.

Applying Neural ODEs directly to detailed chemical mechanisms remains
challenging because of the high dimensionality of the thermochemical state and
the wide range of species concentrations. Major species may have mass
fractions of order unity, whereas the radicals and intermediates that control
ignition and chain branching may be several orders of magnitude less abundant.
A loss function dominated by major species therefore provides inadequate
accuracy for chemically important minor species, and many previous Neural ODE
studies have consequently trained on a selected subset of major species, with
the minor species reconstructed from them. The size of the kinetic system also
increases the training cost and makes it difficult to learn a smooth
representation of stiff chemical dynamics. These difficulties can be addressed
by combining a suitable data transformation with nonlinear dimensionality
reduction. A logarithmic transformation of the species mass fractions,
followed by standardization of each species, maps concentrations spanning many
orders of magnitude onto comparable scales, so that minor species contribute
meaningfully to the training loss \cite{Bansude2023DataDrivenNODE}. An
autoencoder then compresses the transformed state into a latent space with
fewer dimensions and reconstructs the physical state from the latent
coordinates. Unlike linear projection methods, autoencoders can represent
nonlinear correlations among thermochemical variables and approximate the
low-dimensional manifolds occupied by reacting systems. Vijayarangan
\etal\cite{Vijayarangan2024DynamicsInformed} combined an autoencoder with
dynamics-informed training for stiff chemical kinetics and showed that the
latent representation can be smoother and less stiff than the original system.

Data-driven surrogates may nevertheless generate thermochemical states that
violate physical principles, such as negative mass fractions,
$\sum_{i=1}^{N_s}Y_i \neq 1$, or artificial gains and losses of elements.
During extended temporal integration, such violations can accumulate and lead
to drift, physically inadmissible states, or numerical instability.
Physics-informed and physics-constrained learning methods address this by
incorporating governing relationships and conservation principles into
training. Zhang \etal\cite{Zhang2024CRKPINN} embedded chemical and physical
relationships in CRK-PINN for solving combustion kinetics ODEs, and Wang
\etal\cite{Wang2025PhysicalConservation} showed that explicitly enforcing
conservation improves the accuracy and stability of neural network chemistry
surrogates. For Neural ODEs, Kumar \etal\cite{Kumar2025PCNODE} included mass
and elemental conservation in the training objective, and subsequently
introduced Phy-ChemNODE \cite{Kumar2025PhyChemNODE}, which combines an
autoencoder, a Neural ODE, and physical constraints for hydrocarbon fuels.

Despite this progress, several limitations remain. Most studies consider a
single fuel or mechanism, so it is unclear whether one physics-informed latent
framework can represent fuels with different elemental compositions, mechanism
sizes, and kinetic characteristics, such as the nitrogen chemistry of ammonia.
Models are frequently trained and tested on randomly divided samples from
similar trajectories, which measures interpolation accuracy but not the
prediction of complete trajectories at initial temperatures and mixture
compositions excluded from training. Physical constraints are typically
applied to the time-integrated prediction, leaving the consistency of the
learned latent manifold itself less explicitly controlled. Finally, the
computational benefit of these surrogates is rarely examined in terms of how
the learned latent dynamics interact with the choice of numerical solver and
how this benefit changes with mechanism size. Hydrogen, ammonia, and methane
provide a demanding test of these aspects because they represent H/O, H/N/O,
and C/H/O systems with substantially different mechanism sizes, ignition
characteristics, intermediate species, and chemical timescales.

The present study addresses these limitations by developing a Physics-Informed
Autoencoder Neural Ordinary Differential Equation (PI-AE-NODE) framework
together with a complete training and evaluation strategy, spanning data
processing, reduced-order modelling, and numerical performance. Reference
trajectories are logarithmically transformed and standardized so that all
species, including radicals and trace intermediates, are learned together by a
single network rather than reconstructed from a subset of major species. The
encoder, latent Neural ODE, and decoder are trained jointly using trajectory,
direct reconstruction, and latent consistency losses. Elemental conservation,
total mass conservation, and mixture enthalpy consistency are imposed on both
the time-integrated trajectory and the direct autoencoder reconstruction, so
that the latent manifold, and not only its temporal evolution, remains
physically consistent. Separate models with topologically similar
architectures are constructed for H$_2$/air, NH$_3$/air, and CH$_4$/air, with
mechanisms ranging from 9 species and 21 reactions to 52 species and 323
reactions. For H$_2$/air, an otherwise identical data-driven model isolates
the contribution of the physical constraints. The models are evaluated for
major and minor species at interpolated and extrapolated operating conditions
excluded from training, and implicit and explicit ODE solvers are benchmarked
over a range of tolerances to quantify how efficiently the learned latent
dynamics can be integrated as the mechanism grows in size and complexity.

The remainder of this paper is organized as follows.
Section~\ref{sec:methodology} describes the reference data generation,
preprocessing, network architecture, and physics-informed training objective.
Section~\ref{sec:results} presents the results for H$_2$/air, NH$_3$/air, and
CH$_4$/air, including reconstruction accuracy, autoignition trajectories,
conservation behaviour, ignition delay prediction, and computational
performance. Section~\ref{sec:conclusions} summarizes the main findings and
outlines directions for future work.

\section{Methodology}
\label{sec:methodology}


This section describes the complete PI-AE-NODE workflow, from reference data
generation to model evaluation, as summarized in
Fig.~\ref{fig:pi_ae_node_schematic}. Section~\ref{sec:reference_data}
describes the reference detailed kinetic simulations for the three fuel
systems, and Section~\ref{sec:preprocessing} presents the transformation of the
thermochemical state used for training. The encoder, latent Neural ODE, and
decoder are described in Section~\ref{sec:pi_ae_node_architecture}, and the
physics-informed composite objective is formulated in
Section~\ref{sec:composite_objective}. The training configuration is given in
Section~\ref{sec:training_configuration}, and the metrics and procedure used
to assess accuracy and computational performance are defined in later sections.

\subsection{Reference data generation and fuel systems}
\label{sec:reference_data}

Reference thermochemical trajectories were generated using Cantera \cite{cantera} by
simulating homogeneous, adiabatic, constant-pressure, zero-dimensional closed
reactors at atmospheric pressure (\SI{1}{atm}). Three fuel systems were
considered: H$_2$/air, NH$_3$/air, and CH$_4$/air. The H$_2$/air chemistry was
represented using the San Diego mechanism
\cite{Boivin2011HydrogenMechanism,UCSDChemicalMechanisms}. 
For NH$_3$/air, a mechanism containing 24 species and 100 reactions was
derived from the NH$_3$/CH$_4$ mechanism of Okafor
\etal\cite{Okafor2018MethaneAmmonia} by removing carbon-containing species and
their associated reactions. CH$_4$/air combustion was described using GRI-Mech
3.0 \cite{Smith1999GRIMech30}, excluding Ar and the reactions involving Ar. The
resulting mechanisms contain 9, 24, and 52 species and 21, 100, and 323
reactions, respectively.

The homogeneous zero-dimensional reactor was selected because its governing
equations correspond to the chemistry substep in operator-split reacting-flow
simulations. In operator splitting, transport and chemical reaction are
advanced separately, and the local chemistry problem in each computational
cell is equivalent to a homogeneous reactor over the reaction time interval. A
surrogate trained on homogeneous reactor trajectories can therefore replace
the expensive stiff chemistry integration while the existing flow and
transport solvers are retained.

For an adiabatic homogeneous reactor at constant pressure, the species mass
fractions and temperature evolve according to
\begin{equation}
\frac{\mathrm{d}Y_k}{\mathrm{d}t}
=
\frac{W_k\dot{\omega}_k}{\rho},
\qquad
k=1,\ldots,N_s,
\label{eq:species_governing}
\end{equation}
\begin{equation}
\frac{\mathrm{d}T}{\mathrm{d}t}
=
-\frac{1}{\rho c_p}
\sum_{k=1}^{N_s}
h_k W_k\dot{\omega}_k,
\label{eq:temperature_governing}
\end{equation}
where $Y_k$, $W_k$, and $h_k$ are the mass fraction, molecular weight, and
specific enthalpy of species $k$, respectively, $\rho$ is the mixture density,
$c_p$ is the mixture specific heat at constant pressure, and $\dot{\omega}_k$
is the molar production rate of species $k$ obtained from the elementary
reaction rates of the mechanism. Equations~(\ref{eq:species_governing})
and~(\ref{eq:temperature_governing}) form a coupled system of nonlinear
ordinary differential equations, which is numerically stiff because the
characteristic timescales of the elementary reactions differ by several orders
of magnitude.

The initial temperature, $T_0$, and equivalence ratio, $\phi$, were varied to
cover fuel-lean, stoichiometric, and fuel-rich conditions. For H$_2$/air, 15
initial temperatures between 300 and \SI{1500}{K} and 20 equivalence ratios
between 0.1 and 2.0 were used, giving 300 conditions. For NH$_3$/air, 20
initial temperatures between 300 and \SI{2000}{K} and 15 equivalence ratios
between 0.1 and 1.5 were used, also giving 300 conditions. For CH$_4$/air, 19
initial temperatures between 300 and \SI{1700}{K} and 11 equivalence ratios
between 0.1 and 2.0 were used, giving 209 conditions. In all cases, $T_0$ was sampled at \SI{50}{K} intervals from \SI{900}{K}
(H$_2$/air and CH$_4$/air) or \SI{1200}{K} (NH$_3$/air) to the upper limit,
supplemented by lower temperatures of 300 and \SI{600}{K} (and \SI{900}{K} for
NH$_3$/air), so that the model also represents slowly reacting and non-reacting states
For each condition, the reactor equations were integrated from $t=0$ to
$t_{\mathrm{end}}=10^{-2}$, $10^{-1}$, and \SI{e-1}{s} for H$_2$/air,
NH$_3$/air, and CH$_4$/air, respectively, following the procedure of Bansude
\etal\cite{Bansude2023DataDrivenNODE}. 
These intervals capture the induction period, rapid temperature rise,
reactant consumption, product formation, intermediate species evolution, and
approach to the post-ignition state. The reference detailed kinetic model
(DKM) trajectories were computed with the adaptive LSODA solver
using relative and absolute tolerances of $10^{-12}$ and $10^{-15}$,
respectively, to ensure accurate resolution of species present at low
concentrations. Each trajectory was sampled at 300 time points using uniform
fractional time band sampling, in which
to represent the induction, ignition, and post-ignition regimes despite their
substantially different characteristic timescales.
\subsection{Data preprocessing and thermochemical state representation}
\label{sec:preprocessing}

For a mechanism with $N_s$ species, the physical thermochemical state is
$\mathbf{y}=[Y_1,\ldots,Y_{N_s},T]^{\mathrm{T}}$, of dimension $N_s+1$, where
$Y_k$ is the mass fraction of species $k$ and $T$ is the temperature. Because
the mass fractions span many orders of magnitude, training on untransformed
values would cause the loss to be dominated by major species and would
represent radicals and intermediates poorly. Each mass fraction is therefore
logarithmically transformed and standardized, and the temperature is scaled to
the unit interval:
\begin{equation}
x_k=\frac{\ln\left(Y_k+\varepsilon\right)-\mu_k}{\sigma_k},
\quad k=1,\ldots,N_s,
\qquad
x_T=\frac{T-T_{\min}}{T_{\max}-T_{\min}},
\label{eq:forward_transform}
\end{equation}
where $\varepsilon=10^{-10}$ prevents evaluation of the logarithm at zero,
$\mu_k$ and $\sigma_k$ are the mean and standard deviation of
$\ln(Y_k+\varepsilon)$, and $T_{\min}$ and $T_{\max}$ are the temperature
bounds. All statistics are computed separately for each fuel system from its
training data. The resulting network state is
$\mathbf{x}=[x_1,\ldots,x_{N_s},x_T]^{\mathrm{T}}$. No transformation is
applied to time, and the Neural ODE operates on the physical time coordinates
of the reference trajectories.

The decoder output, $\widehat{\mathbf{x}}$, is mapped back to physical
quantities by inverting Eq.~(\ref{eq:forward_transform}):
\begin{align}
\widetilde{Y}_k &= \exp\left(\sigma_k\widehat{x}_k+\mu_k\right),
\label{eq:inverse_species_transform}\\
\widehat{T} &= \widehat{x}_T\left(T_{\max}-T_{\min}\right)+T_{\min}.
\label{eq:inverse_temperature_transform}
\end{align}
The exponential guarantees nonnegative species values, which are then
normalized to satisfy mass conservation:
\begin{equation}
\widehat{Y}_k=\frac{\widetilde{Y}_k}{\sum_{j=1}^{N_s}\widetilde{Y}_j}.
\label{eq:mass_fraction_normalization}
\end{equation}
Both forms are retained during training. The normalized mass fractions,
$\widehat{Y}_k$, are used to evaluate the elemental composition and mixture
enthalpy, whereas the raw values, $\widetilde{Y}_k$, are used in the mass
conservation loss so that it retains a nonzero gradient for the mass conservation constraint
(Section~\ref{sec:composite_objective}).

\subsection{PI-AE-NODE architecture}
\label{sec:pi_ae_node_architecture}

\begin{figure}[htbp]
\centering
\includegraphics[width=\textwidth]{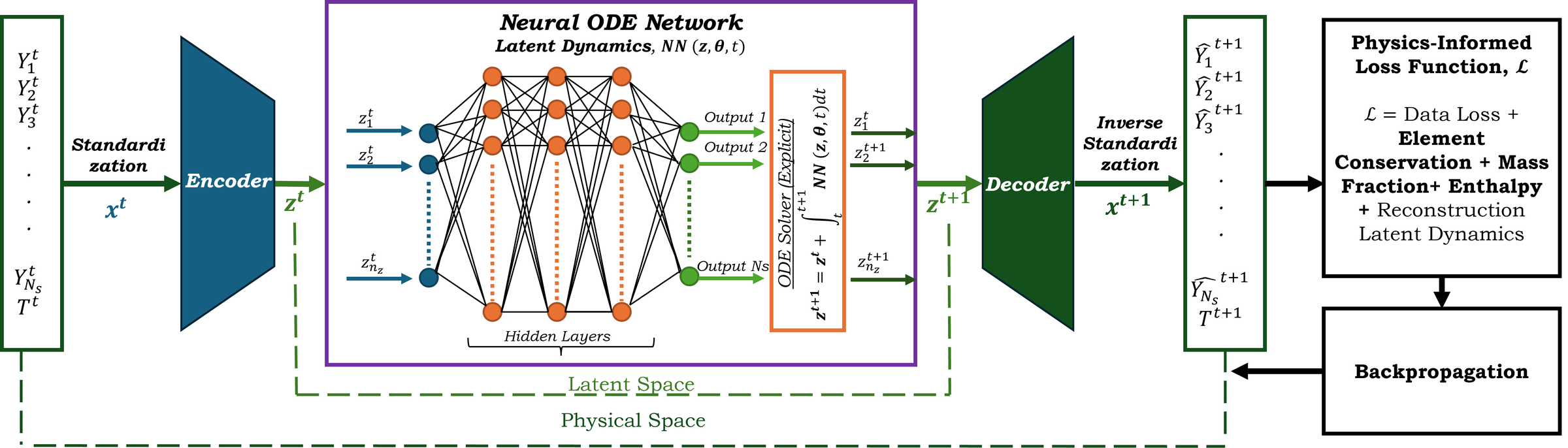}
\caption{Schematic of the PI-AE-NODE framework. The transformed thermochemical
state $\mathbf{x}^t$ is encoded into the latent state $\mathbf{z}^t$, advanced
in time by the latent Neural ODE using an explicit solver, and decoded and
inverse-transformed to the predicted species mass fractions and temperature.
The encoder, Neural ODE, and decoder are trained jointly using a
physics-informed loss that combines data, reconstruction, and latent dynamics
terms with elemental conservation, mass conservation, and enthalpy
consistency.}
\label{fig:pi_ae_node_schematic}
\end{figure}

The PI-AE-NODE surrogate consists of three jointly trained modules: an encoder
$E_{\boldsymbol{\theta}_E}$, a latent Neural ODE $f_{\boldsymbol{\theta}_f}$,
and a decoder $D_{\boldsymbol{\theta}_D}$ (Fig.~\ref{fig:pi_ae_node_schematic}).
The encoder maps the transformed state to the latent coordinate,
$\mathbf{z}=E_{\boldsymbol{\theta}_E}(\mathbf{x})$, and the decoder maps it
back, $\widehat{\mathbf{x}}=D_{\boldsymbol{\theta}_D}(\mathbf{z})$, predicting
the transformed species variables and normalized temperature simultaneously.
Both are fully connected networks with two hidden layers of 128 neurons and
exponential linear unit (ELU) activations, followed by a linear output layer.
The encoder layer sizes are 10--128--128--5, 25--128--128--8, and
53--128--128--10 for H$_2$/air, NH$_3$/air, and CH$_4$/air, respectively, and
the decoders mirror these architectures. The thermochemical state is thus
compressed from 10 to 5, from 25 to 8, and from 53 to 10 variables.
The latent dimensions were selected with reference to previous autoencoder-based
reduced-order models of chemical kinetics
\cite{Vijayarangan2024DynamicsInformed,Kumar2024DeepONets,Kumar2025PhyChemNODE}. Only the input and latent dimensions differ between the three fuel systems.
This common architecture keeps the comparison focused on the training
strategy rather than on fuel-specific network design, and systematic
architecture and hyperparameter optimization, for example using Bayesian
optimization, can be applied subsequently within the same framework. No
bounded activation is applied to the decoder output; instead, positivity and
mass conservation are imposed through
Eqs.~(\ref{eq:inverse_species_transform})
and~(\ref{eq:mass_fraction_normalization}).

The latent state evolves according to the autonomous Neural ODE
\begin{equation}
\frac{\mathrm{d}\mathbf{z}}{\mathrm{d}t}
=
f_{\boldsymbol{\theta}_f}(\mathbf{z}),
\qquad
\mathbf{z}(t_0)=E_{\boldsymbol{\theta}_E}\left[\mathbf{x}(t_0)\right],
\label{eq:latent_ode}
\end{equation}
where $f_{\boldsymbol{\theta}_f}$ is a fully connected network with three
hidden layers of 128 neurons, ELU activations, and a linear output layer, with
input and output dimensions equal to the latent dimension. Time is not
supplied as an input, so the latent dynamics depend only on the current
state. The smooth ELU activation yields a continuously differentiable
right-hand side, which is favourable for adaptive ODE integration. During
training, Eq.~(\ref{eq:latent_ode}) is integrated with the adaptive explicit
fifth-order Dormand--Prince method (\texttt{dopri5}) in \texttt{torchdiffeq},
with absolute and relative tolerances of $10^{-7}$ and $10^{-6}$. Training
with an explicit solver favours latent dynamics that can be advanced
efficiently by explicit integration, which is examined in
Section~\ref{sec:results}. All weights are initialized using Xavier uniform
initialization, and all biases are initialized to zero.

Two prediction pathways are used during training. In the trajectory pathway,
the initial state is encoded, advanced by the Neural ODE, and decoded at each
sampled time. In the reconstruction pathway, each reference state is encoded
and decoded directly, without time integration:
\begin{align}
\widehat{\mathbf{x}}^{\mathrm{tr}}(t_i)
&=
D_{\boldsymbol{\theta}_D}\left[\mathbf{z}(t_i)\right],
\label{eq:trajectory_pathway}\\
\widehat{\mathbf{x}}^{\mathrm{rec}}(t_i)
&=
D_{\boldsymbol{\theta}_D}\left\{
E_{\boldsymbol{\theta}_E}\left[\mathbf{x}^{\mathrm{DKM}}(t_i)\right]
\right\}.
\label{eq:reconstruction_pathway}
\end{align}
The trajectory pathway contains the combined errors of the encoder, latent
dynamics, numerical integration, and decoder, whereas the reconstruction
pathway isolates the information lost through dimensionality reduction.
Applying the physical constraints to both pathways
(Section~\ref{sec:composite_objective}) encourages the learned latent
manifold, as well as its temporal evolution, to remain consistent with
physically admissible thermochemical states.

\subsection{Physics-informed composite objective}
\label{sec:composite_objective}
The framework is trained to reproduce all species, including radicals and
intermediates several orders of magnitude less abundant than the major
species. This increases the difficulty of optimization but is necessary
because minor species govern ignition, chain branching, heat release, and
pollutant formation.

All loss components use the mean absolute error, $\operatorname{MAE}(\mathbf{a},
\mathbf{b})$, defined as the mean of the elementwise absolute differences
between two tensors of identical dimensions.
The data losses compare the model predictions with the DKM reference in the
transformed space of Eq.~(\ref{eq:forward_transform}):
\begin{equation}
\begin{aligned}
\mathcal{L}_{Y}^{\mathrm{tr}} &=
\operatorname{MAE}\left(\widehat{\mathbf{x}}_{Y}^{\mathrm{tr}},
\mathbf{x}_{Y}^{\mathrm{DKM}}\right),
&\qquad
\mathcal{L}_{T}^{\mathrm{tr}} &=
\operatorname{MAE}\left(\widehat{x}_{T}^{\mathrm{tr}},
x_{T}^{\mathrm{DKM}}\right),
\\
\mathcal{L}_{\mathrm{rec}} &=
\operatorname{MAE}\left(\widehat{\mathbf{x}}^{\mathrm{rec}},
\mathbf{x}^{\mathrm{DKM}}\right),
&\qquad
\mathcal{L}_{\mathrm{lat}} &=
\operatorname{MAE}\left(\mathbf{z}^{\mathrm{ODE}},
E_{\boldsymbol{\theta}_E}\left[\mathbf{x}^{\mathrm{DKM}}\right]\right).
\end{aligned}
\label{eq:data_losses}
\end{equation}
Here, $\mathcal{L}_{Y}^{\mathrm{tr}}$ and $\mathcal{L}_{T}^{\mathrm{tr}}$ are
the species and temperature trajectory losses, $\mathcal{L}_{\mathrm{rec}}$ is
the direct reconstruction loss over all transformed variables, and
$\mathcal{L}_{\mathrm{lat}}$ aligns the integrated latent state with the
encoding of the corresponding reference state at each sampled time.

The physical-consistency losses are evaluated in physical space after the
inverse transformation and are applied to both the trajectory and
reconstruction pathways, denoted by $p\in\{\mathrm{tr},\mathrm{rec}\}$. The
elemental mass fraction of element $e$ for a composition $\mathbf{Y}$ is
\begin{equation}
Z_e(\mathbf{Y})=\sum_{k=1}^{N_s}\frac{a_{e,k}W_e}{W_k}Y_k,
\label{eq:elemental_mass_fraction}
\end{equation}
where $W_e$ is the atomic weight of element $e$ and $a_{e,k}$ is the number of
atoms of $e$ in species $k$. The elemental, mass, and enthalpy losses are
\begin{equation}
\begin{aligned}
\mathcal{L}_{\mathrm{elem}}^{p} &=
\sum_{e}w_e\,\operatorname{MAE}\left[Z_e\left(\widehat{\mathbf{Y}}^{p}\right),
Z_e\left(\mathbf{Y}^{\mathrm{DKM}}\right)\right],
\\
\mathcal{L}_{\mathrm{mass}}^{p} &=
\operatorname{MAE}\left(\sum_{k=1}^{N_s}\widetilde{Y}_k^{p},\,1\right),
\\
\mathcal{L}_{h}^{p} &=
\operatorname{MAE}\left(\overline{h}^{p},\,\overline{h}^{\mathrm{DKM}}\right),
\qquad
\overline{h}=\frac{h_{\mathrm{mix}}-h_{\min}}{h_{\max}-h_{\min}},
\qquad
h_{\mathrm{mix}}=\sum_{k=1}^{N_s}Y_k h_k(T).
\end{aligned}
\label{eq:physics_losses}
\end{equation}
The element set is $\{\mathrm{H,O,N}\}$ for H$_2$/air and NH$_3$/air and
$\{\mathrm{H,O,N,C}\}$ for CH$_4$/air. The weights $w_e$ are identical for both pathways and are chosen according to
the expected magnitudes of the elemental errors, so that elements with small
absolute errors are not neglected during optimization \cite{Kumar2025PCNODE}. 
The elemental and enthalpy losses use the normalized mass fractions
$\widehat{Y}_k$ of Eq.~(\ref{eq:mass_fraction_normalization}), whereas the mass
loss uses the raw values $\widetilde{Y}_k$ of
Eq.~(\ref{eq:inverse_species_transform}). Because normalization enforces
$\sum_k\widehat{Y}_k=1$ exactly, only the raw values provide a nonzero
gradient that drives the decoder towards mass-conserving outputs. The species
enthalpies $h_k(T)$ are evaluated using differentiable NASA thermodynamic
polynomials, and $h_{\min}$ and $h_{\max}$ are obtained from the initial
states of the training data. The enthalpy target is the DKM mixture enthalpy
at each sampled time; for the adiabatic, constant-pressure reference reactor,
this equals the initial enthalpy to within the integration tolerance. The complete training objective is
\begin{equation}
\mathcal{L}_{\mathrm{total}}
=
\lambda_Y\mathcal{L}_{Y}^{\mathrm{tr}}
+\lambda_T\mathcal{L}_{T}^{\mathrm{tr}}
+\lambda_{\mathrm{rec}}\mathcal{L}_{\mathrm{rec}}
+\lambda_{\mathrm{lat}}\mathcal{L}_{\mathrm{lat}}
+\sum_{p\in\{\mathrm{tr},\mathrm{rec}\}}
\left(
\lambda_{\mathrm{elem}}\mathcal{L}_{\mathrm{elem}}^{p}
+\lambda_{\mathrm{mass}}\mathcal{L}_{\mathrm{mass}}^{p}
+\lambda_{h}\mathcal{L}_{h}^{p}
\right),
\label{eq:total_training_loss}
\end{equation}
where the coefficients $\lambda$ control the relative contributions of the
individual terms. 
For the data-driven model (DD-AE-NODE) used in the H$_2$/air comparison, the
elemental, mass, and enthalpy terms are excluded from backpropagation but are
still computed as diagnostic quantities for direct comparison with
PI-AE-NODE.




\subsection{Model training configuration}
\label{sec:training_configuration}

The same training procedure was applied to the three fuel systems, with the input and latent dimensions adjusted according to the corresponding kinetic mechanism.
The model parameters were optimized using the Adam algorithm with an initial
learning rate of $2\times10^{-3}$ for 2000 epochs. Each mini-batch consisted
of one complete trajectory of 300 sampled time points, so that every update
included the induction, ignition, and post-ignition regimes. The order of the
training trajectories was randomly shuffled in each epoch. The learning rate
was reduced using a plateau-based scheduler whenever the mean training loss did not improve for 25 epochs, down to a
minimum of $10^{-6}$. 
\subsection{Evaluation metrics and computational benchmarking}
\label{sec:evaluation_metrics}

Accuracy is assessed separately for the direct reconstruction and the
time-integrated trajectories. For each operating condition, the species and
temperature errors are defined as
\begin{equation}
\varepsilon_Y=\frac{1}{N_tN_s}\sum_{i=1}^{N_t}\sum_{k=1}^{N_s}
\left|\widehat{Y}_k(t_i)-Y_k^{\mathrm{DKM}}(t_i)\right|,
\qquad
\varepsilon_T=\frac{1}{N_t}\sum_{i=1}^{N_t}
\left|\widehat{T}(t_i)-T^{\mathrm{DKM}}(t_i)\right|,
\label{eq:error_metrics}
\end{equation}
where $N_t=300$ is the number of sampled times. The ignition delay time,
$\tau_{\mathrm{ign}}$, is defined as the first time at which the temperature
reaches $T_0+\SI{400}{K}$, and is reported as the mean relative error with
respect to the DKM over all igniting conditions. 

Computational performance is quantified by the speedup
$S=t_{\mathrm{DKM}}/t_{\mathrm{PI\text{-}AE\text{-}NODE}}$, where
$t_{\mathrm{DKM}}$ and $t_{\mathrm{PI\text{-}AE\text{-}NODE}}$ are the
wall-clock times required to integrate all evaluated trajectories. The DKM reference is integrated with the implicit BDF method. PI-AE-NODE is
integrated with the same implicit BDF method at the loosest tolerances that
preserve its accuracy, and with the explicit RK45 and RK23 methods over a
range of absolute and relative tolerances.

\section{Results and discussion}
\label{sec:results}

The PI-AE-NODE framework is evaluated for H$_2$/air, NH$_3$/air, and
CH$_4$/air using the metrics defined in Section~\ref{sec:evaluation_metrics},
with the DKM solution as the reference. For each fuel, direct autoencoder
reconstruction is assessed first to quantify the information retained in the
latent representation, followed by time-integrated trajectories, which also
include the errors of the learned latent dynamics and numerical integration.
For H$_2$/air, PI-AE-NODE is additionally compared with a data-driven
counterpart (DD-AE-NODE) trained without the elemental, mass, and enthalpy
losses, to isolate the effect of the physical constraints.

\subsection{H$_2$/air system}
\label{sec:h2_results}

The H$_2$/air model compresses the ten-dimensional thermochemical state, nine
species mass fractions and temperature, into five latent variables.
PI-AE-NODE and DD-AE-NODE share the same architecture, training data,
preprocessing, and data losses, and differ only in the physical constraints
included in the training objective.
\subsubsection{Training convergence and physical consistency}
\label{sec:h2_training_convergence}

\begin{figure}[!htbp]
\centering
\includegraphics[width=\textwidth]{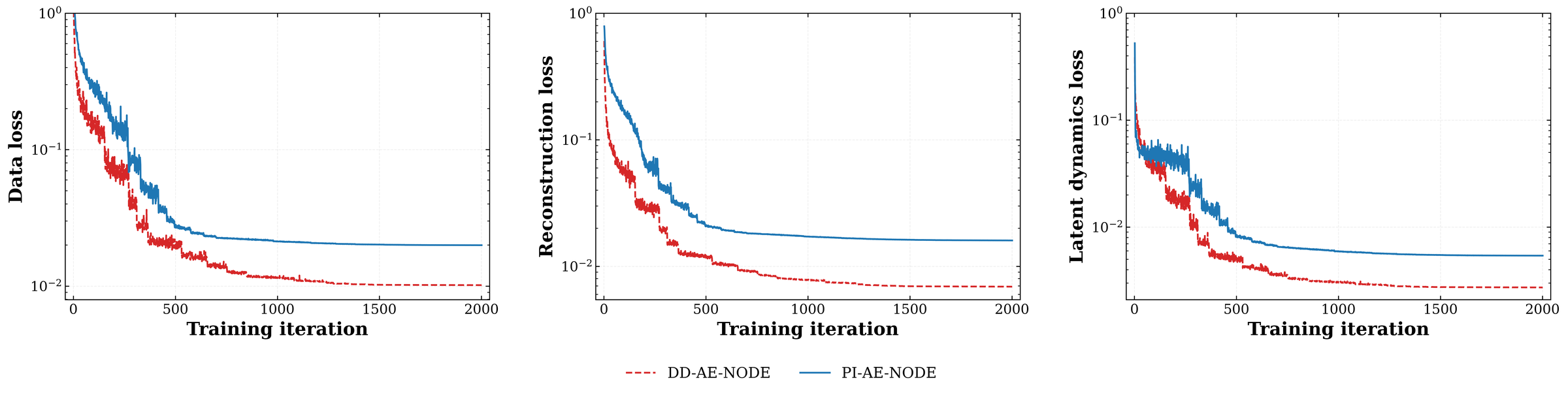}
\caption{Training loss histories of DD-AE-NODE and PI-AE-NODE for H$_2$/air: (a) combined species and temperature data loss, (b) reconstruction loss, and (c) latent dynamics loss. All vertical axes are shown on logarithmic scales.}
\label{fig:h2_common_training_losses}
\end{figure}

Because PI-AE-NODE and DD-AE-NODE minimize different objectives, their
training histories are compared component by component. For the loss terms
common to both models (Fig.~\ref{fig:h2_common_training_losses}), all losses
decrease rapidly during the initial iterations and then approach nearly
constant values, indicating stable convergence. DD-AE-NODE reaches moderately
lower values because its optimization is directed only towards data fitting,
reconstruction, and latent dynamics, whereas PI-AE-NODE must balance these
objectives against the additional physical constraints. Its moderately higher
common losses should therefore be interpreted together with the corresponding
physical residuals.

\begin{figure}[!htbp]
\centering
\includegraphics[width=\textwidth]{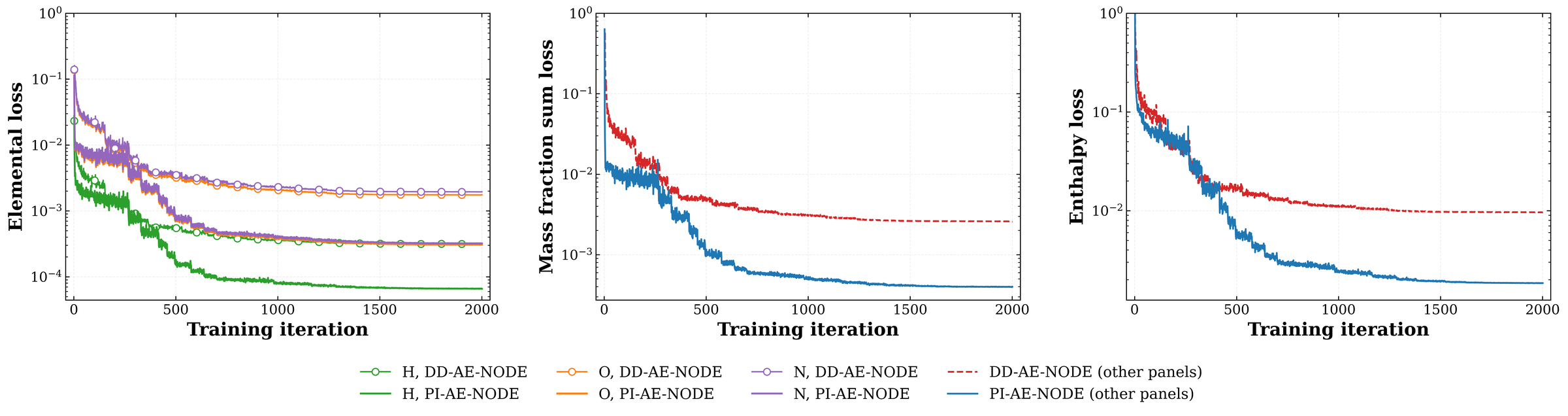}
\caption{\textit{Trajectory pathway:} physical consistency losses of the
time-integrated H$_2$/air predictions of DD-AE-NODE and PI-AE-NODE:
(a) H, O, and N elemental conservation, (b) mass conservation, and
(c) enthalpy consistency. For DD-AE-NODE, these losses are diagnostic only.
All vertical axes are logarithmic.}
\label{fig:h2_trajectory_physics_losses}
\end{figure}

\begin{figure}[!htbp]
\centering
\includegraphics[width=\textwidth]{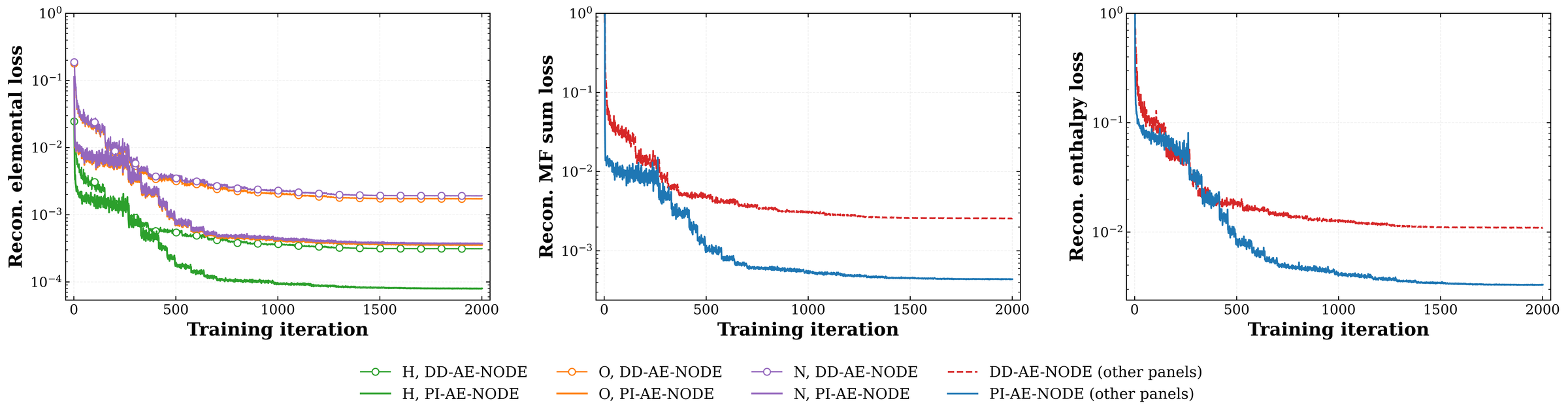}
\caption{\textit{Reconstruction pathway:} physical consistency losses of the
directly reconstructed H$_2$/air states (without time integration) of
DD-AE-NODE and PI-AE-NODE: (a) H, O, and N elemental conservation, (b) mass
conservation, and (c) enthalpy consistency. For DD-AE-NODE, these losses are
diagnostic only. All vertical axes are logarithmic.}
\label{fig:h2_reconstruction_physics_losses}
\end{figure}

The physical residuals are shown for the two prediction pathways of
Section~\ref{sec:pi_ae_node_architecture}.
Figure~\ref{fig:h2_trajectory_physics_losses} presents the elemental, mass,
and enthalpy residuals of the time-integrated trajectories, which include the
errors of the latent dynamics and numerical integration.
Figure~\ref{fig:h2_reconstruction_physics_losses} presents the same residuals
for the directly reconstructed states, obtained by encoding and decoding the
reference states without time integration. The two figures therefore have the
same layout but isolate different error sources. In both pathways, the
physical residuals of both models decrease during training, indicating that
data fitting alone partially improves physical consistency. However, the
residuals of DD-AE-NODE plateau at higher values, whereas PI-AE-NODE reduces
them by up to approximately an order of magnitude. The improvement in the
reconstruction pathway shows that the constraints shape the learned latent
manifold itself, not only its temporal evolution. The advantage of PI-AE-NODE
therefore lies not in a lower data loss but in a better balance between data
accuracy and the physical relationships among the species mass fractions and
temperature.


\subsubsection{Autoencoder reconstruction performance}
\label{sec:h2_reconstruction}

\begin{figure}[!htbp]
\centering
\includegraphics[width=\textwidth]{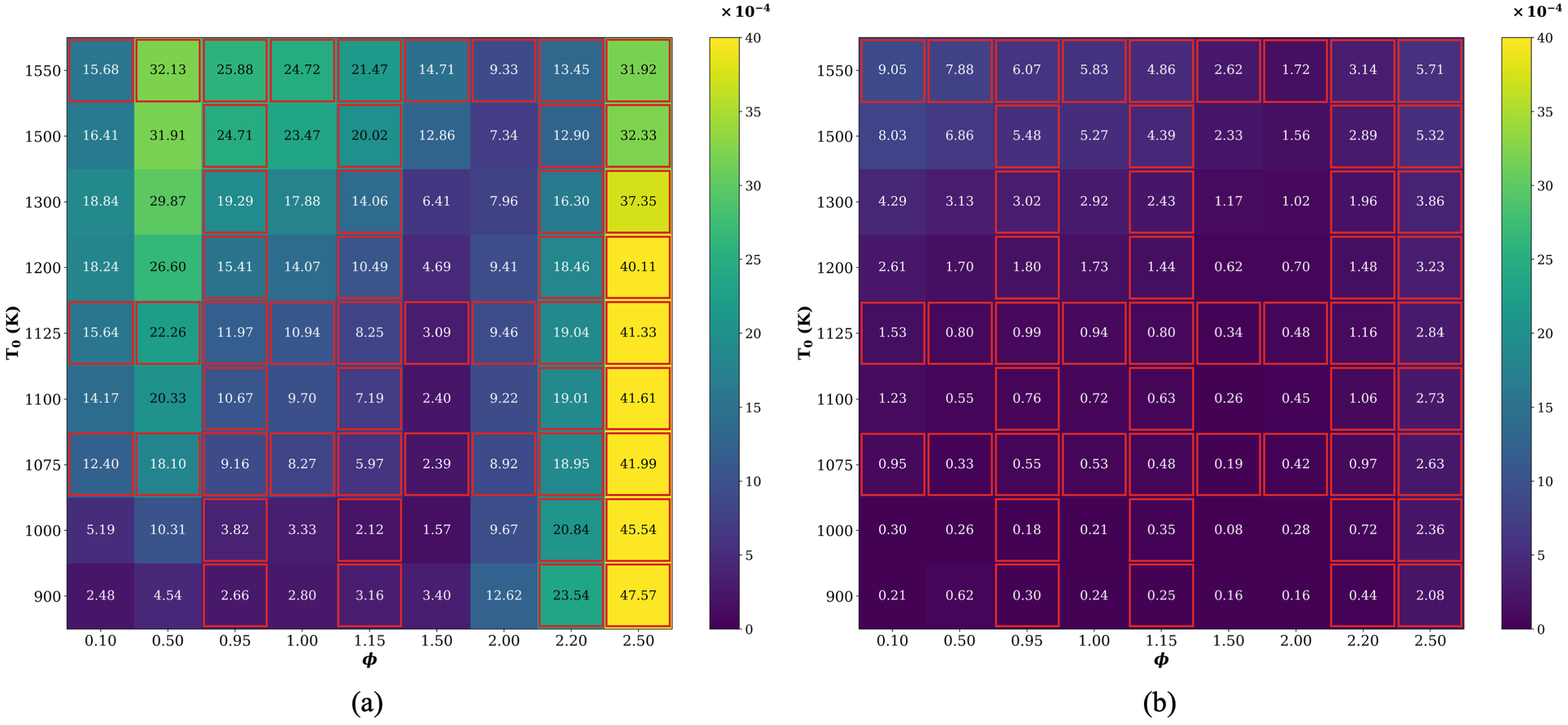}
\caption{Species reconstruction error $\varepsilon_Y$
(Eq.~\ref{eq:error_metrics}) for H$_2$/air: (a) data-driven and (b)
physics-informed autoencoders. Boxed cells denote conditions excluded from
training.}
\label{fig:h2_species_rec}
\end{figure}

\begin{figure}[!htbp]
\centering
\includegraphics[width=\textwidth]{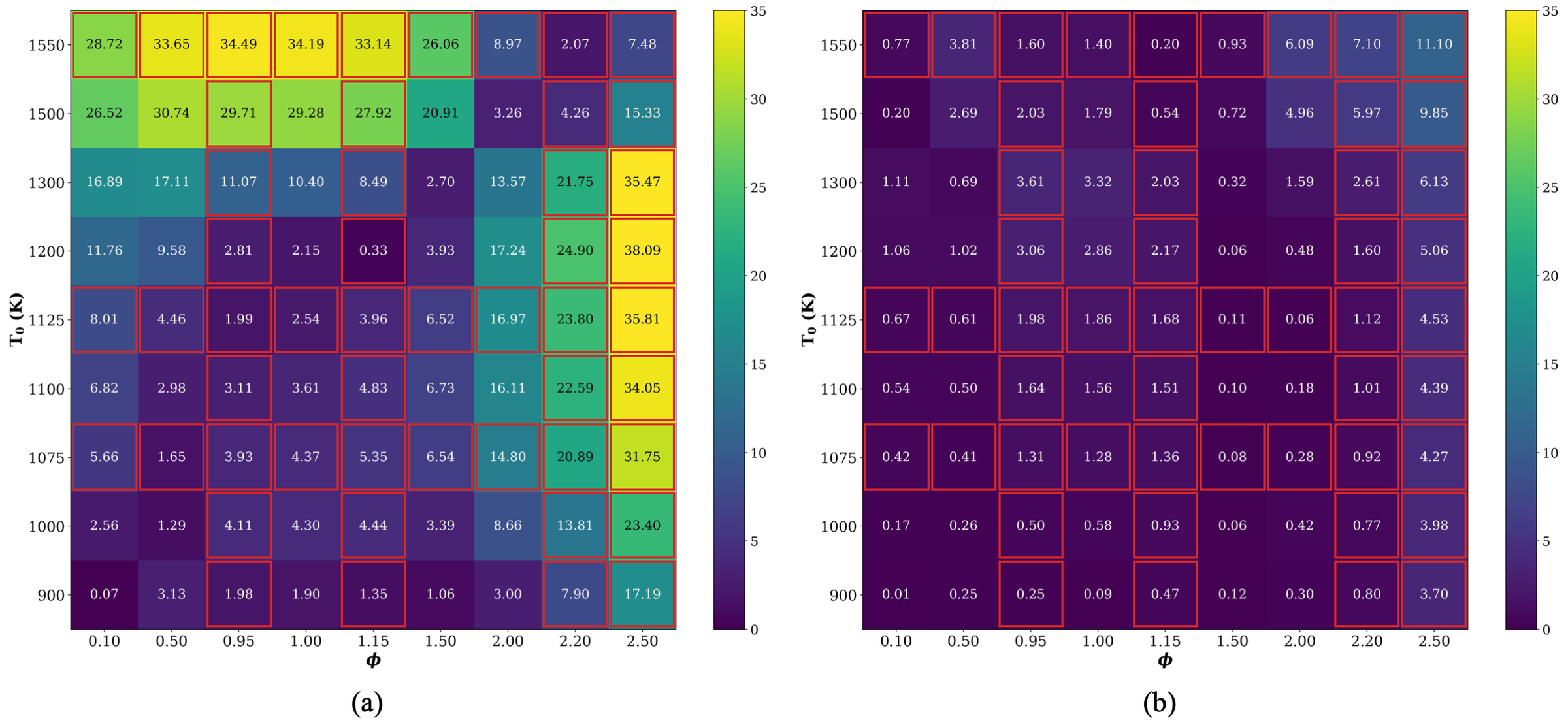}
\caption{Temperature reconstruction error $\varepsilon_T$ (K)
(Eq.~\ref{eq:error_metrics}) for H$_2$/air: (a) data-driven and (b)
physics-informed autoencoders. Boxed cells denote conditions excluded from
training.}\label{fig:h2_temperature_rec}
\end{figure}

Direct reconstruction isolates the quality of the learned latent
representation, because neither the latent Neural ODE nor time integration is
involved. Since the elemental, mass, and enthalpy losses are also applied to
this pathway in PI-AE-NODE, any improvement relative to DD-AE-NODE reflects
the ability of the constrained autoencoder to preserve the coupled
relationships among the species mass fractions and temperature. The evaluation
grid in Figs.~\ref{fig:h2_species_rec} and~\ref{fig:h2_temperature_rec}
combines training conditions with unseen intermediate values and extrapolated values of initial temperature and
equivalence ratio.

The species reconstruction errors, $\varepsilon_Y$
(Eq.~\ref{eq:error_metrics}), of the data-driven autoencoder lie between
$10^{-4}$ and $10^{-3}$ over most of the domain and increase to approximately
$4.8\times10^{-3}$ under rich conditions, particularly at $\phi=2.5$
(Fig.~\ref{fig:h2_species_rec}). The physics-informed autoencoder reduces
these errors to approximately $10^{-5}$--$10^{-4}$ over most of the domain,
with a maximum of $9.1\times10^{-4}$ at the extrapolated temperature
$T_0=\SI{1550}{K}$. This improvement is maintained at both the unseen
intermediate and the extrapolated conditions, indicating that elemental
conservation, mass conservation, and enthalpy consistency produce a latent
representation that more accurately preserves the coupled species
composition.

The corresponding temperature errors, $\varepsilon_T$, are shown in
Fig.~\ref{fig:h2_temperature_rec}. The data-driven autoencoder exhibits errors
of up to approximately \SI{38}{K}, with the largest deviations at elevated
initial temperatures and rich conditions. The physics-informed autoencoder
reduces the maximum error to approximately \SI{11.1}{K} and keeps it below
\SI{5}{K} for most conditions. Moderate increases occur only at the most
strongly extrapolated rich conditions, where the errors nevertheless remain
well below those of the data-driven model. The simultaneous reduction of the
species and temperature errors indicates that the physical constraints improve
the thermochemical consistency of the reconstructed state as a whole, rather
than the accuracy of an individual variable.

\subsubsection{In-domain autoignition trajectories}
\label{sec:h2_in_domain}

\begin{figure}[!htbp]
\centering
\includegraphics[width=\textwidth]{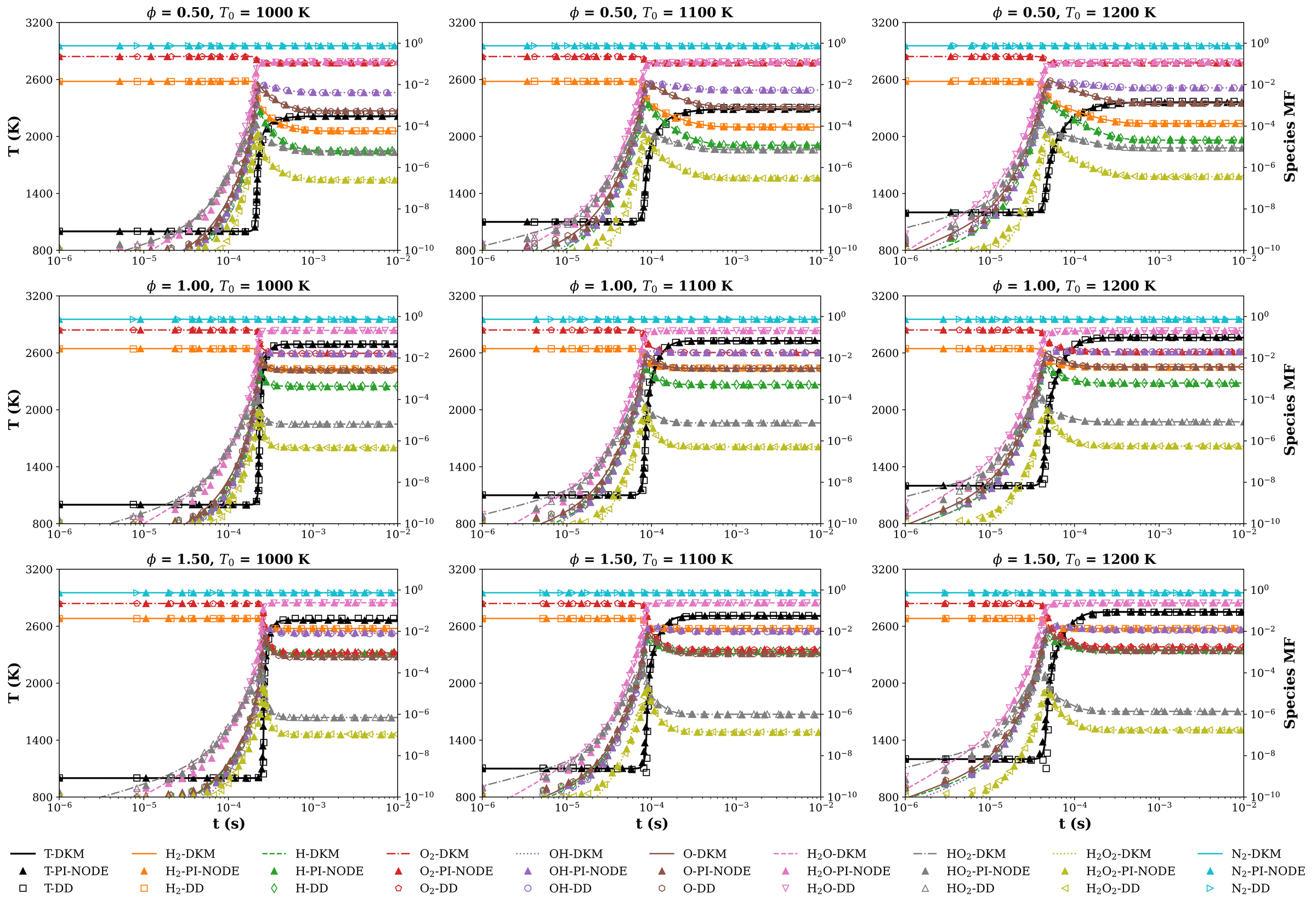}
\caption{Temperature and species mass fraction trajectories predicted by DKM,
PI-AE-NODE, and DD-AE-NODE for H$_2$/air at the training conditions
$\phi=0.5$, 1.0, and 1.5 and $T_0=1000$, 1100, and \SI{1200}{K}.}
\label{fig:h2_in_domain}
\end{figure}

The complete PI-AE-NODE model was evaluated over the \SI{0.01}{s} prediction
interval at training conditions spanning lean, stoichiometric, and rich
mixtures ($\phi=0.5$, 1.0, and 1.5) and $T_0=1000$, 1100, and \SI{1200}{K}
(Fig.~\ref{fig:h2_in_domain}). PI-AE-NODE closely follows the DKM temperature
profiles for all nine conditions and reproduces the approximately fivefold
decrease in ignition delay as $T_0$ increases from 1000 to \SI{1200}{K}. The
predicted species profiles capture H$_2$ and O$_2$ consumption, H$_2$O
formation, the transition to the post-ignition state, and the dependence of
the final composition on equivalence ratio. DD-AE-NODE exhibits larger local
deviations near ignition, where the temperature and species change most
rapidly, indicating that the physical constraints improve the temporal
evolution of the decoded state in addition to its direct reconstruction.

\begin{figure}[!htbp]
\centering
\includegraphics[width=0.85\textwidth]{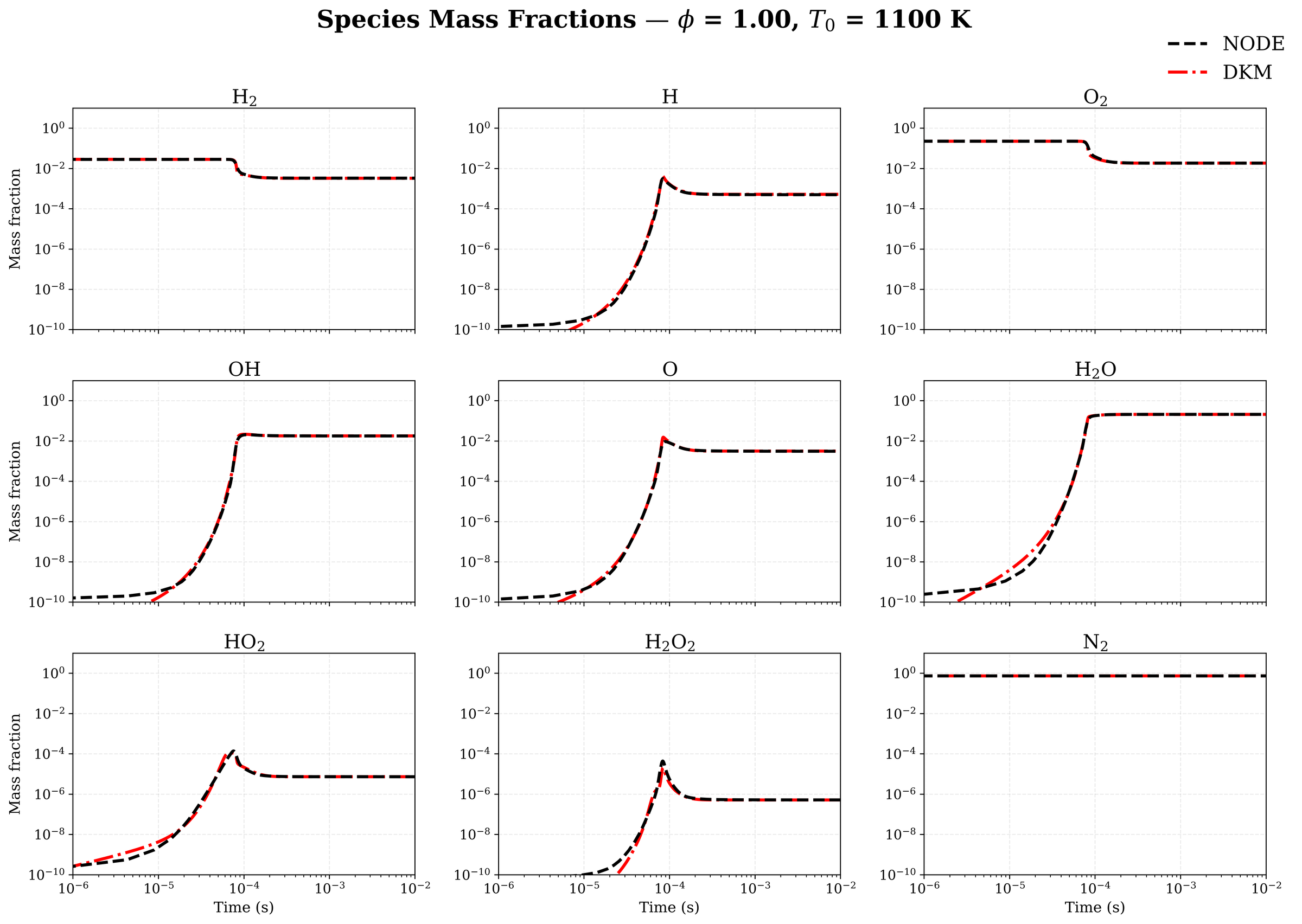}
\caption{Mass fractions of all nine species predicted by DKM and PI-AE-NODE
for H$_2$/air at $\phi=1.0$ and $T_0=\SI{1100}{K}$.}
\label{fig:h2_in_domain_species}
\end{figure}

Figure~\ref{fig:h2_in_domain_species} shows all nine species at $\phi=1.0$
and $T_0=\SI{1100}{K}$, which is representative of the agreement obtained at
the other conditions. PI-AE-NODE accurately captures the consumption of H$_2$
and O$_2$, the formation of H$_2$O, and the nearly constant N$_2$ mass
fraction. It also reproduces the gradual accumulation and subsequent
consumption of HO$_2$ and H$_2$O$_2$ during the induction period and the rapid
formation of H, O, and OH near ignition. Minor differences appear only where
the intermediate mass fractions are extremely low, early in the induction
period, while their peak locations, decay behaviour, and post-ignition values
agree closely with the DKM. The learned latent dynamics therefore retain both
the main heat-release process and the radical chemistry governing H$_2$/air
autoignition.

\subsubsection{Unseen-condition autoignition trajectories}
\label{sec:h2_unseen}

\begin{figure}[!htpb]
\centering
\includegraphics[width=\textwidth]{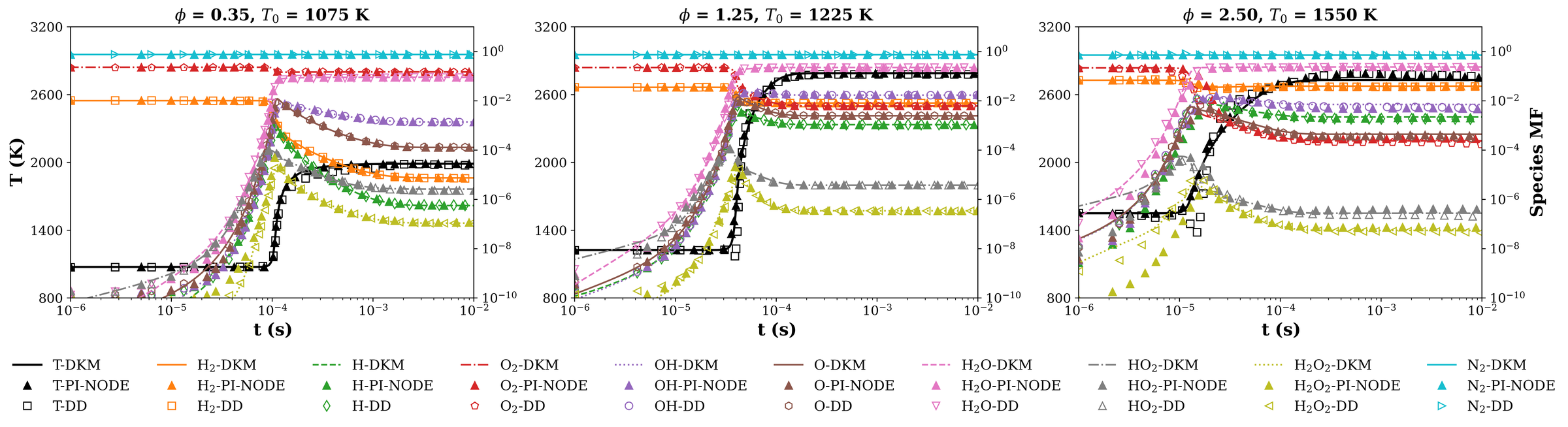}
\caption{Temperature and species mass fraction trajectories predicted by DKM,
PI-AE-NODE, and DD-AE-NODE for H$_2$/air at the unseen conditions
$(\phi,T_0)=(0.35,\SI{1075}{K})$ and $(1.25,\SI{1225}{K})$ (interpolated) and
$(2.50,\SI{1550}{K})$ (extrapolated).}
\label{fig:h2_unseen}
\end{figure}

Generalization was evaluated at three conditions excluded from training
(Fig.~\ref{fig:h2_unseen}). The conditions $(\phi,T_0)=(0.35,\SI{1075}{K})$
and $(1.25,\SI{1225}{K})$ lie between the training values of both parameters,
whereas $(2.50,\SI{1550}{K})$ extends beyond the training ranges of both
equivalence ratio and initial temperature. At the two interpolated
conditions, PI-AE-NODE closely follows the DKM temperature and species
profiles through the induction, ignition, and post-ignition periods,
reproducing H$_2$ and O$_2$ consumption, H$_2$O formation, radical evolution,
and ignition timing. At the extrapolated condition, PI-AE-NODE retains the
ignition timing and post-ignition state, with larger deviations confined
mainly to low-concentration species during the induction period. DD-AE-NODE
shows comparatively larger deviations near ignition and for low-concentration
species at all three conditions, and these become most pronounced under
extrapolation.

\begin{figure}[!htbp]
\centering
\includegraphics[width=0.85\textwidth]{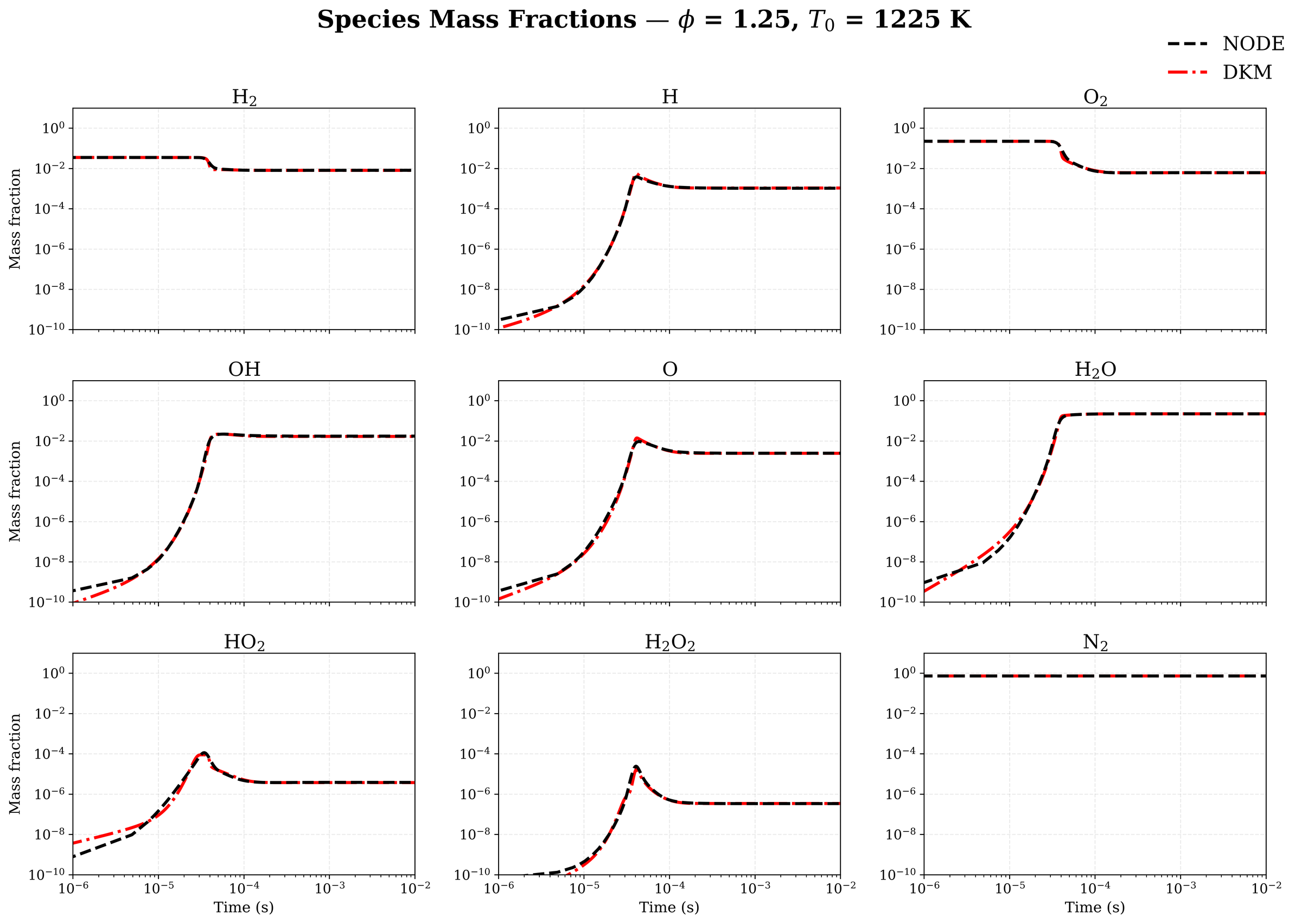}
\caption{Mass fractions of all nine species predicted by DKM and PI-AE-NODE
for H$_2$/air at the unseen condition $\phi=1.25$ and $T_0=\SI{1225}{K}$.}
\label{fig:h2_unseen_species}
\end{figure}

Figure~\ref{fig:h2_unseen_species} shows all nine species at the interpolated
condition $\phi=1.25$ and $T_0=\SI{1225}{K}$. PI-AE-NODE closely reproduces the
DKM profiles of both stable species and reactive intermediates. Small
differences occur only early in the induction period, where the mass
fractions are extremely low, while peak locations, decay behaviour, and
post-ignition values agree closely with the DKM. The model therefore retains
the essential chemical features of the H$_2$/air mechanism at conditions
excluded from training.

\subsubsection{Elemental and mass conservation}
\label{sec:h2_conservation}

\begin{figure}[!htbp]
\centering
\includegraphics[width=\textwidth]{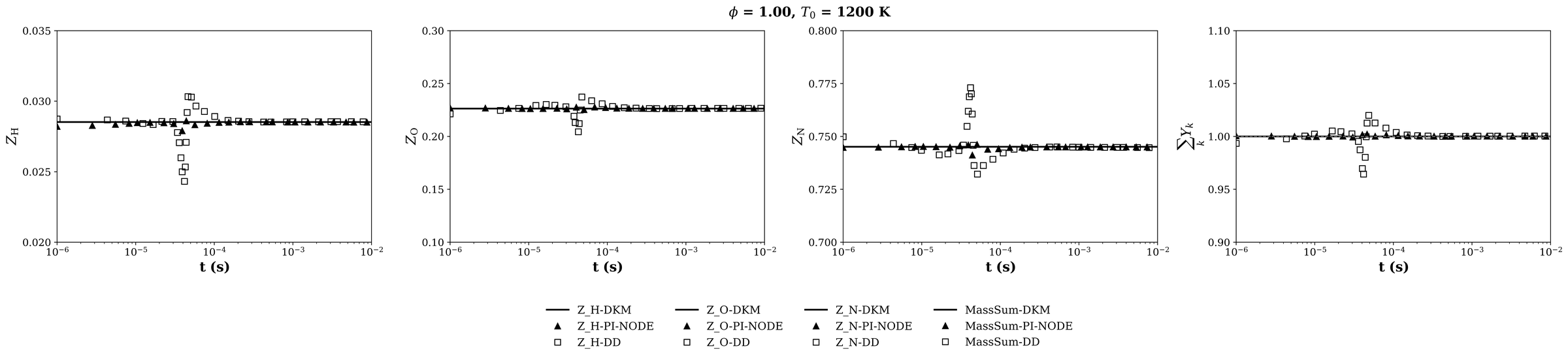}
\caption{Hydrogen, oxygen, and nitrogen elemental mass fractions and total species mass fraction sum obtained from DKM, PI-AE-NODE, and DD-AE-NODE for H$_2$/air at $\phi=1.00$ and $T_0=1200$ K. The elemental reference values are calculated from the initial composition, and the reference value of the species mass fraction sum is unity.}
\label{fig:h2_conservation_in}
\end{figure}

\begin{figure}[!htbp]
\centering
\includegraphics[width=\textwidth]{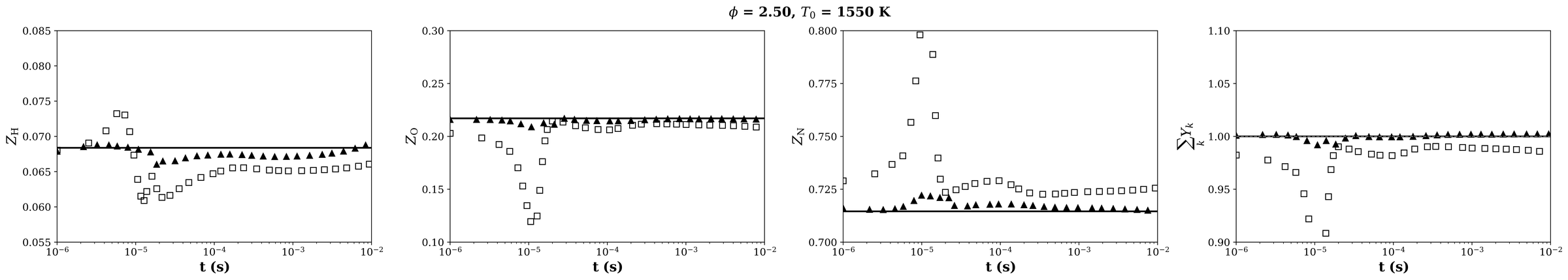}
\caption{As Fig.~\ref{fig:h2_conservation_in}, at the extrapolated condition
$\phi=2.50$ and $T_0=\SI{1550}{K}$.}
\label{fig:h2_conservation_ex}
\end{figure}

Elemental conservation and the sum of the species mass fractions provide
stricter measures of physical consistency than comparisons of individual
species trajectories. Figures~\ref{fig:h2_conservation_in}
and~\ref{fig:h2_conservation_ex} compare the H, O, and N elemental mass
fractions, $Z_e$ (Eq.~\ref{eq:elemental_mass_fraction}), and the species mass
fraction sum, $\sum_k\widetilde{Y}_k$, predicted by PI-AE-NODE and DD-AE-NODE.
The exact references are the elemental mass fractions of the initial
composition, which remain constant in a closed reactor, and
$\sum_k Y_k=1$. The DKM trajectories, computed with strict solver tolerances,
deviate negligibly from these values and are shown as the numerical
benchmark.

At the training condition $\phi=1.0$ and $T_0=\SI{1200}{K}$
(Fig.~\ref{fig:h2_conservation_in}), PI-AE-NODE keeps all elemental mass
fractions and the mass fraction sum close to their reference values
throughout the prediction interval, including the rapid ignition period.
DD-AE-NODE exhibits pronounced transient deviations of several percent near
ignition, particularly in the H and N elemental mass fractions and the mass
fraction sum. Although these deviations decay after ignition, they show that
agreement in individual species profiles does not guarantee conservation of
elemental composition and total mass.

At the extrapolated condition $\phi=2.50$ and $T_0=\SI{1550}{K}$
(Fig.~\ref{fig:h2_conservation_ex}), both models exhibit larger conservation
errors. The deviations of DD-AE-NODE are substantially larger and persist
after ignition, whereas PI-AE-NODE remains close to the exact references and
the DKM, with only a small offset in the H elemental mass fraction. Applying
the elemental and mass conservation penalties to both the trajectory and
reconstruction pathways therefore improves physical consistency, including
under extrapolation.

\subsubsection{Ignition delay prediction}
\label{sec:h2_ignition_delay}

\begin{figure}[!htbp]
\centering
\includegraphics[width=0.75\textwidth]{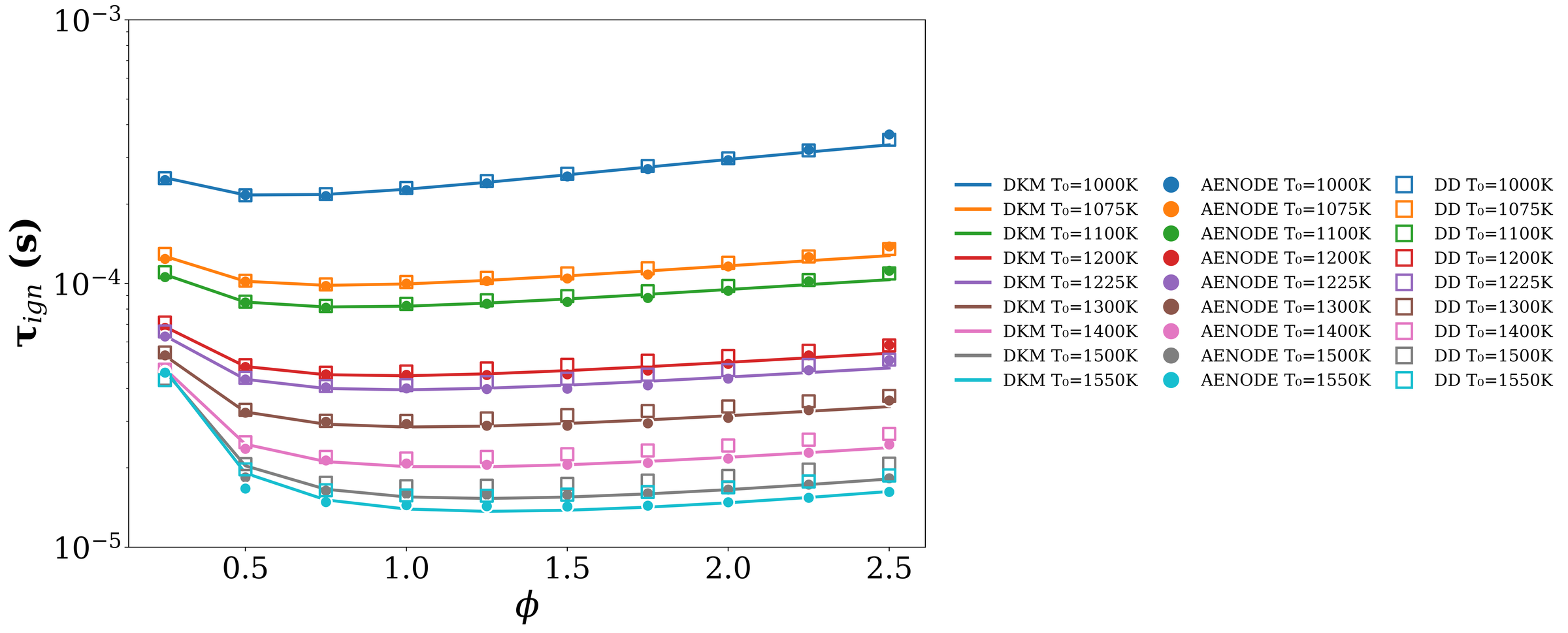}
\caption{Ignition delays of H$_2$/air predicted by DKM (line), PI-AE-NODE (circle), and
DD-AE-NODE (square) as a function of equivalence ratio for different initial
temperatures.}
\label{fig:h2_ignition_delay}
\end{figure}

Figure~\ref{fig:h2_ignition_delay} compares the ignition delays, defined in
Section~\ref{sec:evaluation_metrics}, predicted by DKM, PI-AE-NODE, and
DD-AE-NODE over the investigated ranges of initial temperature and
equivalence ratio, including unseen and extrapolated conditions. The ignition
delay decreases by more than an order of magnitude as $T_0$ increases from
1000 to \SI{1550}{K} and varies only weakly with $\phi$, reaching a shallow
minimum under lean to stoichiometric conditions. PI-AE-NODE reproduces these
trends closely across the complete range, with an overall mean relative error
of 2.20\%. DD-AE-NODE yields a mean relative error of 6.19\%, with the largest
deviations at intermediate and high initial temperatures, where it
systematically overpredicts the ignition delay. The physical constraints
therefore reduce the ignition delay error by a factor of almost three and
improve prediction consistency across the operating range.

\subsubsection{Computational performance and accuracy trade-off}
\label{sec:h2_computational_performance}

The computational performance of PI-AE-NODE for H$_2$/air was evaluated
following Section~\ref{sec:evaluation_metrics}, using the solvers and
tolerances listed in Table~\ref{tab:h2_computational_performance}. To provide
a fair baseline, the DKM tolerances were chosen as the loosest that preserve
its accuracy, including that of low-concentration species. PI-AE-NODE, in
contrast, integrates latent variables derived from standardized inputs, which
vary over a much narrower range than the physical mass fractions. Its
tolerances are therefore not directly comparable to those of the DKM and were
scaled to maintain a comparable level of relative accuracy. Further details of
this tolerance selection and benchmarking procedure can be found in
\cite{Bansude2023DataDrivenNODE,Bansude2023PMSR}.

\begin{table}[!htbp]
\centering
\caption{Computational performance and ignition delay accuracy of PI-AE-NODE for H$_2$/air using different ODE solvers and tolerance settings.}
\label{tab:h2_computational_performance}

{\fontfamily{ptm}\fontsize{9}{10.35}\selectfont
\renewcommand{\arraystretch}{1.05}
\setlength{\tabcolsep}{3pt}

\begin{tabular}{
@{}
>{\centering\arraybackslash}p{0.8cm}
>{\centering\arraybackslash}p{2.2cm}
>{\centering\arraybackslash}p{1.3cm}
>{\centering\arraybackslash}p{1.8cm}
>{\centering\arraybackslash}p{1.2cm}
>{\centering\arraybackslash}p{1.5cm}
>{\centering\arraybackslash}p{1.8cm}
@{}}
\toprule
\textbf{Case}
& \textbf{Chemistry model}
& \textbf{Solver}
& \textbf{ATOL}
& \textbf{RTOL}
& \textbf{Speedup}
& \textbf{Ignition delay error (\%)} \\
\midrule
Base & DKM        & BDF  & $10^{-12}$ ($10^{-6}$) & $10^{-8}$ & --   & --   \\
1    & PI-AE-NODE & BDF  & $10^{-5}$               & $10^{-4}$ & 1.56 & 2.20 \\
2    & PI-AE-NODE & BDF  & $10^{-4}$               & $10^{-3}$ & 2.39 & 2.20 \\
3    & PI-AE-NODE & RK45 & $10^{-6}$               & $10^{-5}$ & 3.03 & 2.20 \\
4    & PI-AE-NODE & RK45 & $10^{-5}$               & $10^{-4}$ & 4.22 & 2.18 \\
5    & PI-AE-NODE & RK45 & $10^{-4}$               & $10^{-3}$ & 5.11 & 2.42 \\
6    & PI-AE-NODE & RK23 & $10^{-4}$               & $10^{-3}$ & 5.69 & 2.20 \\
\bottomrule
\end{tabular}
}
\end{table}

With the implicit BDF solver, PI-AE-NODE achieves speedups of $1.56\times$
and $2.39\times$ (Cases 1 and 2). At the same tolerances, the explicit RK45
solver is $2.1$--$2.7$ times faster than BDF, and its speedup increases from
$3.03\times$ to $5.11\times$ as the tolerances are relaxed (Cases 3--5). The
highest speedup of $5.69\times$ is obtained with RK23 (Case 6). The mean
ignition delay error remains between 2.18\% and 2.42\% in all cases,
indicating that it is governed by the learned model rather than by the
integration tolerance, so Case 6 provides the most favourable balance between
efficiency and accuracy. Implicit
integration offers no advantage for the latent Neural ODE, and the explicit
solvers become faster as the tolerances are relaxed, indicating that their
step size is governed by accuracy rather than by numerical stability. This
behaviour is consistent with training the latent dynamics using an explicit
solver. The modest absolute speedup for H$_2$/air reflects the small size of
the mechanism, for which evaluating the detailed source terms costs about the
same as evaluating the neural network.
\subsection{NH$_3$/air system}
\label{sec:nh3_results}

The NH$_3$/air mechanism contains 24 species and 100 reactions, giving a
25-dimensional thermochemical state that is represented by eight latent
variables. Compared with hydrogen, ammonia oxidation involves additional
coupled pathways through nitrogen-containing species such as NH$_2$, NH, HNO,
NO, and N$_2$O, and spans a much broader range of ignition timescales. This
system therefore tests whether PI-AE-NODE can retain the coupled radical and
nitrogen chemistry after substantial dimensionality reduction. The evaluation
follows the same sequence as for H$_2$/air: reconstruction, in-domain and
unseen-condition trajectories, ignition delay, and computational performance.
\subsubsection{Autoencoder reconstruction performance}
\label{sec:nh3_reconstruction}

\begin{figure}[!htbp]
    \centering
    \begin{subfigure}[t]{0.51\textwidth}
        \centering
        \includegraphics[width=\linewidth]{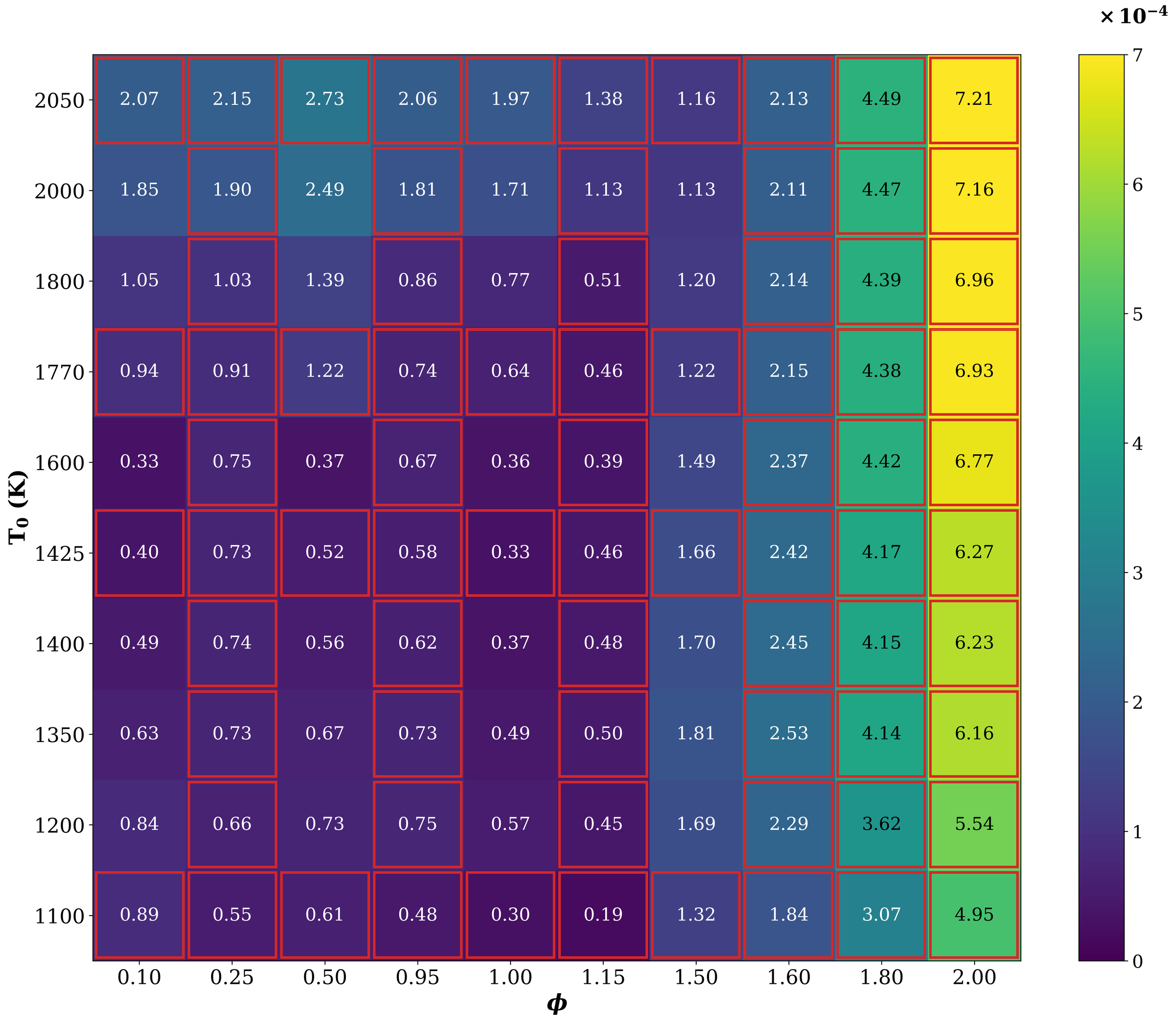}
        \caption{}
        \label{fig:nh3_species_rec}
    \end{subfigure}
    \hfill
    \begin{subfigure}[t]{0.48\textwidth}
        \centering
        \includegraphics[width=\linewidth]{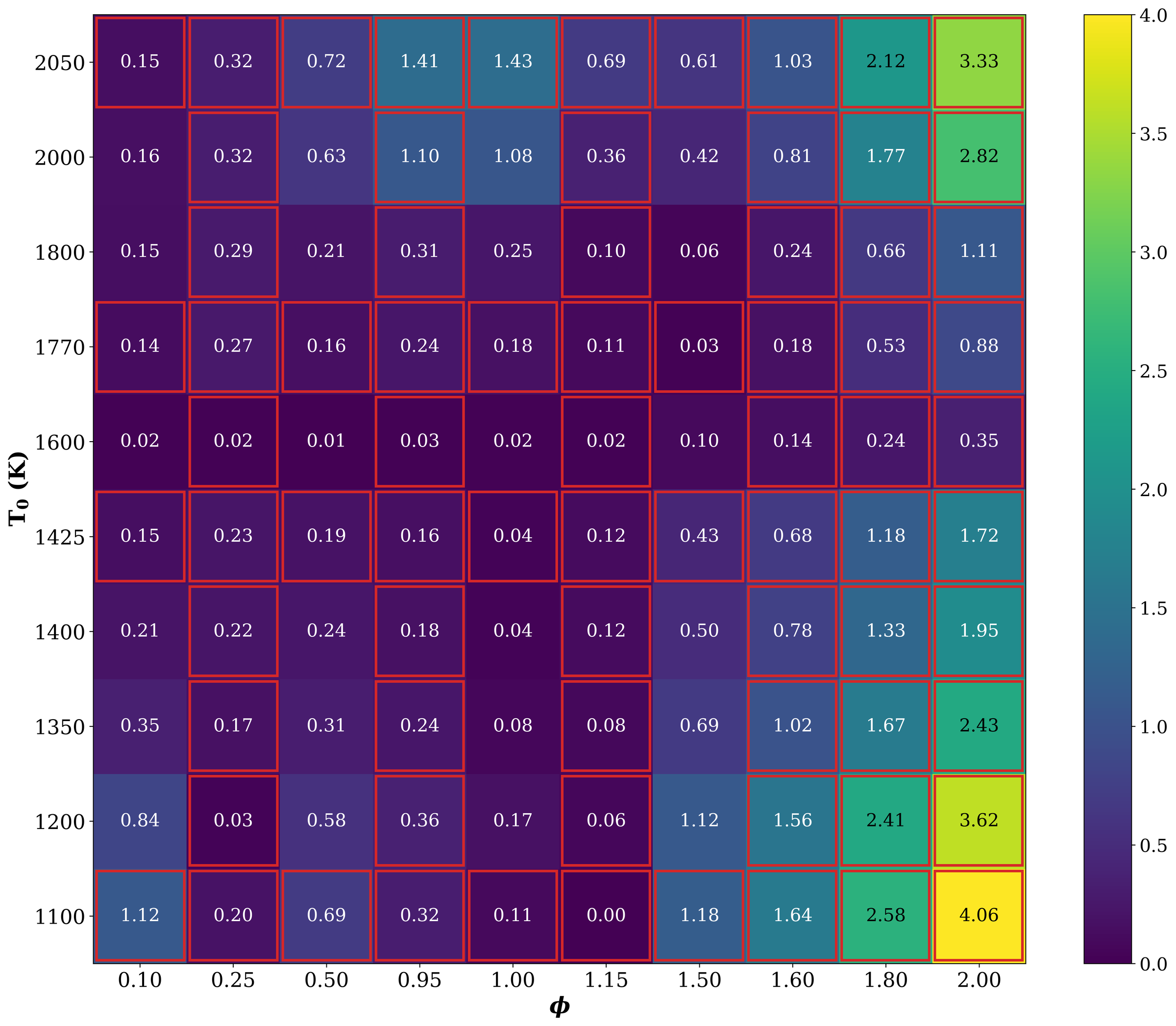}
        \caption{}
        \label{fig:nh3_temperature_rec}
    \end{subfigure}
    \caption{Reconstruction errors of the NH$_3$/air autoencoder: (a) species
error $\varepsilon_Y$ and (b) temperature error $\varepsilon_T$ (K), as
defined in Eq.~(\ref{eq:error_metrics}). Boxed cells denote conditions
excluded from training.}
    \label{fig:nh3_reconstruction}
\end{figure}

The reconstruction errors in Figs.~\ref{fig:nh3_species_rec}
and~\ref{fig:nh3_temperature_rec} are evaluated on a grid that combines
training conditions with unseen intermediate values and extrapolated values. The species reconstruction
error remains below approximately $2.7\times10^{-4}$ for $\phi\le1.6$, with
many conditions below $10^{-4}$ (Fig.~\ref{fig:nh3_species_rec}). It increases
at the extrapolated rich conditions $\phi=1.8$ and 2.0, reaching approximately
$4.5\times10^{-4}$ and $7.2\times10^{-4}$, respectively, but remains below
$10^{-3}$ throughout. The temperature reconstruction error remains below
\SI{1}{K} for most lean, stoichiometric, and moderately rich conditions
(Fig.~\ref{fig:nh3_temperature_rec}). Larger errors occur mainly at
$\phi=1.8$ and 2.0, reaching a maximum of \SI{4.06}{K} at $\phi=2.0$ and
$T_0=\SI{1100}{K}$, where both parameters lie outside the training grid. The
eight-dimensional latent representation therefore preserves the principal
compositional and thermal information of the 25-dimensional state, with
accuracy degrading gradually rather than abruptly under extrapolation.

Figure~\ref{fig:nh3_temperature_rec} presents the corresponding temperature reconstruction errors. The maximum error is approximately 4.06 K and occurs at $\phi=2.0$ and $T_0=1100$ K. The error remains below 1 K for a large proportion of the evaluated conditions, particularly at lean, stoichiometric, and moderately rich compositions. Larger errors occur mainly at $\phi=1.8$ and 2.0, including conditions outside the training domain. Even at these extrapolated conditions, the temperature error remains below approximately 4.1 K. The species and temperature results demonstrate that the autoencoder provides an accurate and physically consistent reduced representation of the NH$_3$/air thermochemical state before temporal evolution through the latent Neural ODE.

\subsubsection{In-domain autoignition trajectories}
\label{sec:nh3_in_domain}

\begin{figure}[!htbp]
\centering
\includegraphics[width=\textwidth]{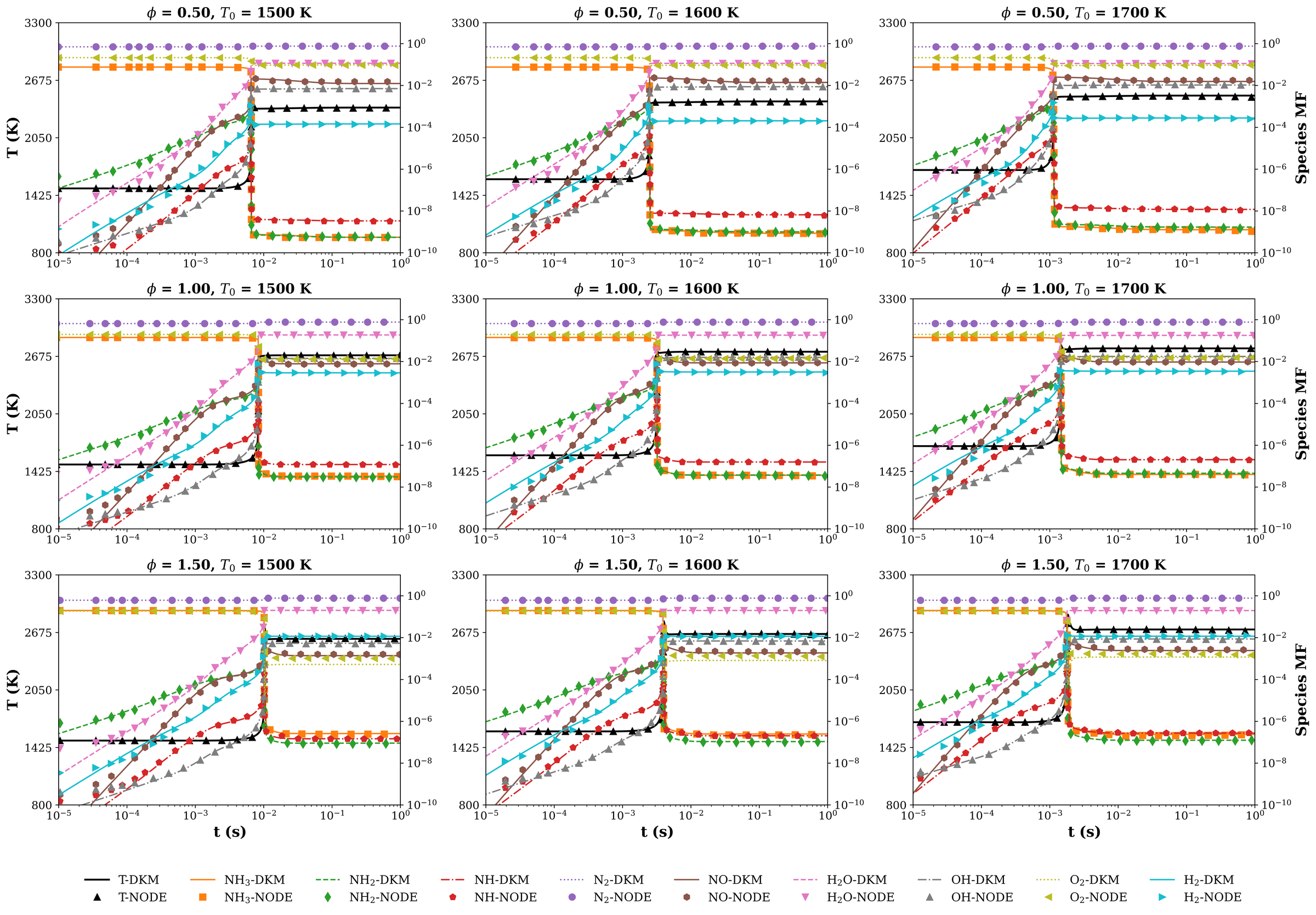}
\caption{Temperature and species mass fraction trajectories predicted by DKM
and PI-AE-NODE for NH$_3$/air at the training conditions $\phi=0.5$, 1.0, and
1.5 and $T_0=1500$, 1600, and \SI{1700}{K}.}
\label{fig:nh3_in_domain}
\end{figure}

The complete PI-AE-NODE model was evaluated over the \SI{0.1}{s} prediction
interval at the training conditions $\phi=0.5$, 1.0, and 1.5 and $T_0=1500$,
1600, and \SI{1700}{K} (Fig.~\ref{fig:nh3_in_domain}). For all nine
conditions, PI-AE-NODE reproduces the induction period, rapid temperature
rise, and approach to the post-ignition state, as well as the approximately
sixfold decrease in ignition delay from 1500 to \SI{1700}{K} and the
dependence of the final temperature and composition on equivalence ratio. The
predicted trajectories capture the consumption of NH$_3$ and O$_2$, the
formation of H$_2$O and N$_2$, and the evolution of H$_2$, O, OH, NH$_2$, NH,
and NO during induction and ignition. Because these intermediates change
rapidly near ignition and span several orders of magnitude, their agreement
with the DKM is a sensitive test of ignition timing, and indicates that the
learned latent dynamics retain the coupled heat-release, radical, and
nitrogen pathways of NH$_3$/air autoignition.

\begin{figure}[!htbp]
\centering
\includegraphics[width=\textwidth,height=0.85\textheight,keepaspectratio]{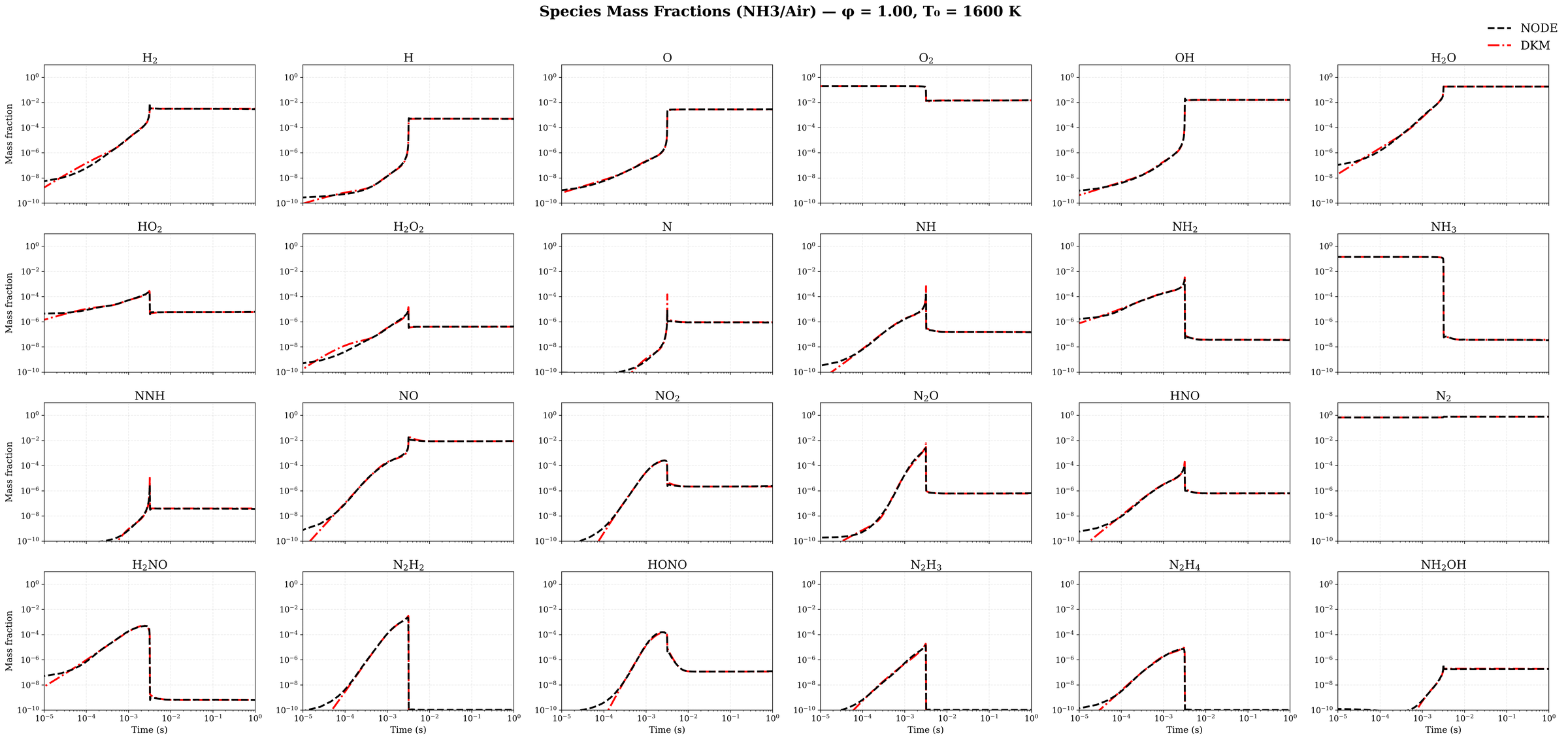}
\caption{Mass fractions of all 24 species predicted by DKM and PI-AE-NODE for
NH$_3$/air at $\phi=1.0$ and $T_0=\SI{1600}{K}$.}
\label{fig:nh3_in_domain_species}
\end{figure}

Figure~\ref{fig:nh3_in_domain_species} shows all 24 species at $\phi=1.0$
and $T_0=\SI{1600}{K}$. In addition to the major species and the radicals H,
O, and OH, PI-AE-NODE captures the formation and consumption of the
nitrogen-containing intermediates N, NH, NH$_2$, NNH, NO, NO$_2$, N$_2$O,
HNO, H$_2$NO, N$_2$H$_2$, HONO, N$_2$H$_3$, N$_2$H$_4$, and NH$_2$OH, as well
as the low-concentration species HO$_2$ and H$_2$O$_2$. Minor local
deviations occur for some trace species near their narrow maxima, where steep
temporal gradients amplify small differences in ignition timing, but the
profile shapes, peak locations, and post-ignition values remain in close
agreement with the DKM.

Previous Neural ODE modeling of ammonia combustion considered seven selected
species: NH$_3$, O$_2$, H$_2$, OH, H$_2$O, N$_2$, and NO
\cite{Saito2023AmmoniaNODE}. The present framework instead reconstructs and
evolves all 24 species within an eight-dimensional latent space, which is a
more demanding test because both the major thermochemical evolution and the
minor nitrogen chemistry must be preserved through the compressed
representation.

\subsubsection{Unseen-condition autoignition trajectories}
\label{sec:nh3_unseen}

\begin{figure}[!htbp]
\centering
\includegraphics[width=\textwidth]{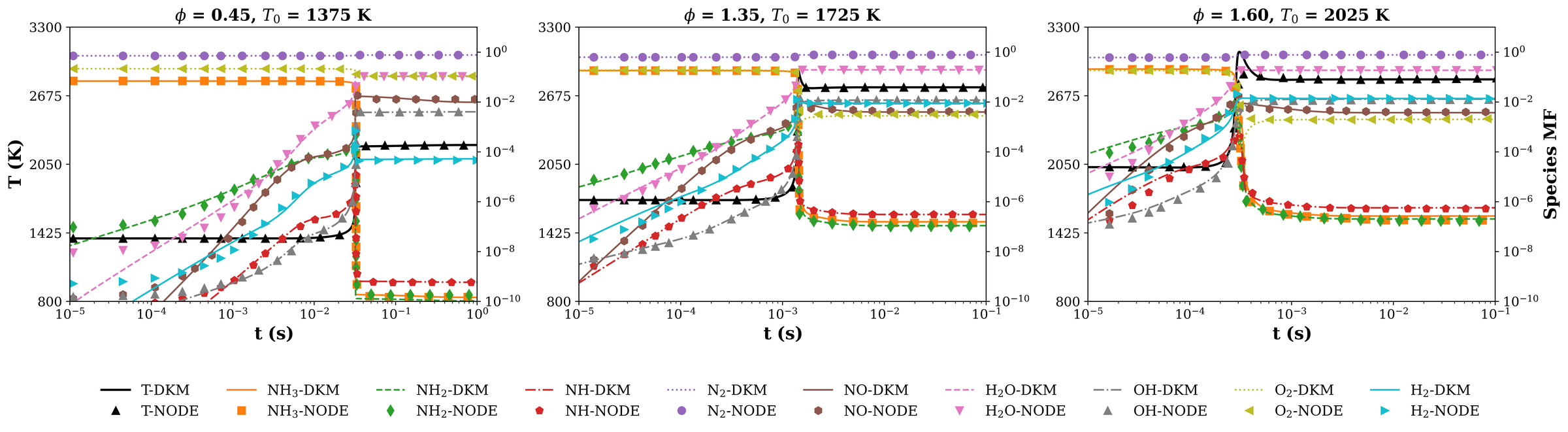}
\caption{Temperature and species mass fraction trajectories predicted by DKM
and PI-AE-NODE for NH$_3$/air at the unseen conditions
$(\phi,T_0)=(0.45,\SI{1375}{K})$ and $(1.35,\SI{1725}{K})$ (interpolated) and
$(1.60,\SI{2025}{K})$ (extrapolated).}
\label{fig:nh3_unseen}
\end{figure}

Generalization was evaluated at three conditions excluded from training
(Fig.~\ref{fig:nh3_unseen}). The conditions $(\phi,T_0)=(0.45,\SI{1375}{K})$
and $(1.35,\SI{1725}{K})$ lie between the training values, whereas
$(1.60,\SI{2025}{K})$ extends beyond the training ranges of both equivalence
ratio and initial temperature. At the lean interpolated condition,
PI-AE-NODE reproduces the relatively long induction period, the rapid
temperature rise, and the approach to the post-ignition state. Close
agreement in ignition timing, temperature evolution, and species composition
is also obtained at $\phi=1.35$ and $T_0=\SI{1725}{K}$. At the extrapolated
condition, PI-AE-NODE captures the much shorter ignition timescale, with a
small local temperature overshoot and minor species deviations near ignition,
while the post-ignition temperature and composition remain consistent with
the DKM.

\begin{figure}[!htbp]
\centering
\includegraphics[width=\textwidth,height=0.85\textheight,keepaspectratio]{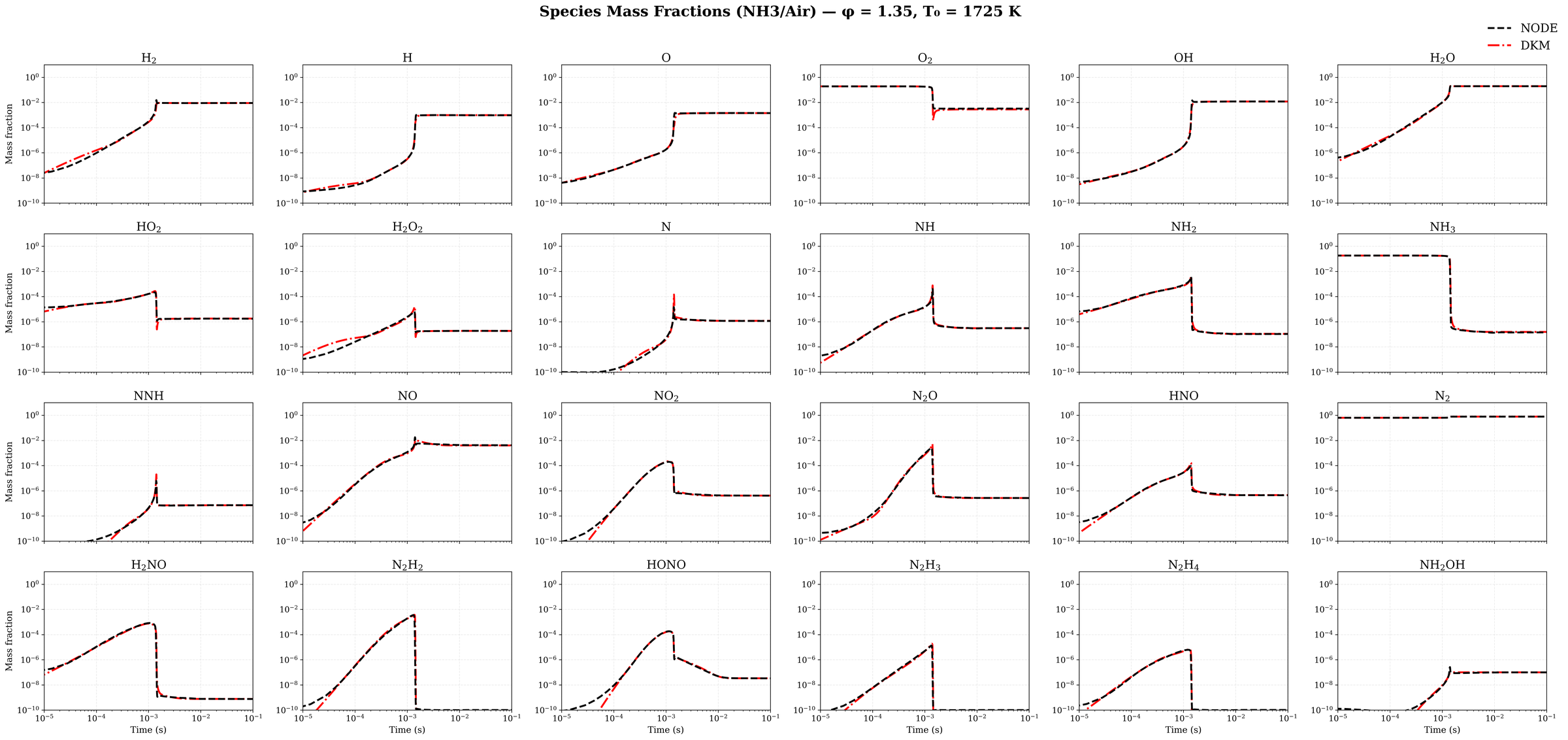}
\caption{Mass fractions of all 24 species predicted by DKM and PI-AE-NODE for
NH$_3$/air at the unseen condition $\phi=1.35$ and $T_0=\SI{1725}{K}$.}
\label{fig:nh3_unseen_species}
\end{figure}

Figure~\ref{fig:nh3_unseen_species} shows all 24 species at the interpolated
condition $\phi=1.35$ and $T_0=\SI{1725}{K}$. PI-AE-NODE accurately
reproduces the major species, the radicals H, O, OH, HO$_2$, and H$_2$O$_2$,
and the same set of nitrogen-containing intermediates as in
Fig.~\ref{fig:nh3_in_domain_species}. Minor deviations occur for a few trace
species early in the induction period and near their narrow peaks, but they
have little influence on the major species, temperature evolution, or
ignition timing. The learned latent dynamics therefore interpolate reliably
across initial temperature and mixture composition and retain useful
predictive capability under moderate extrapolation.

\subsubsection{Ignition delay prediction}
\label{sec:nh3_ignition_delay}

\begin{figure}[!htbp]
\centering
\includegraphics[width=0.75\textwidth]{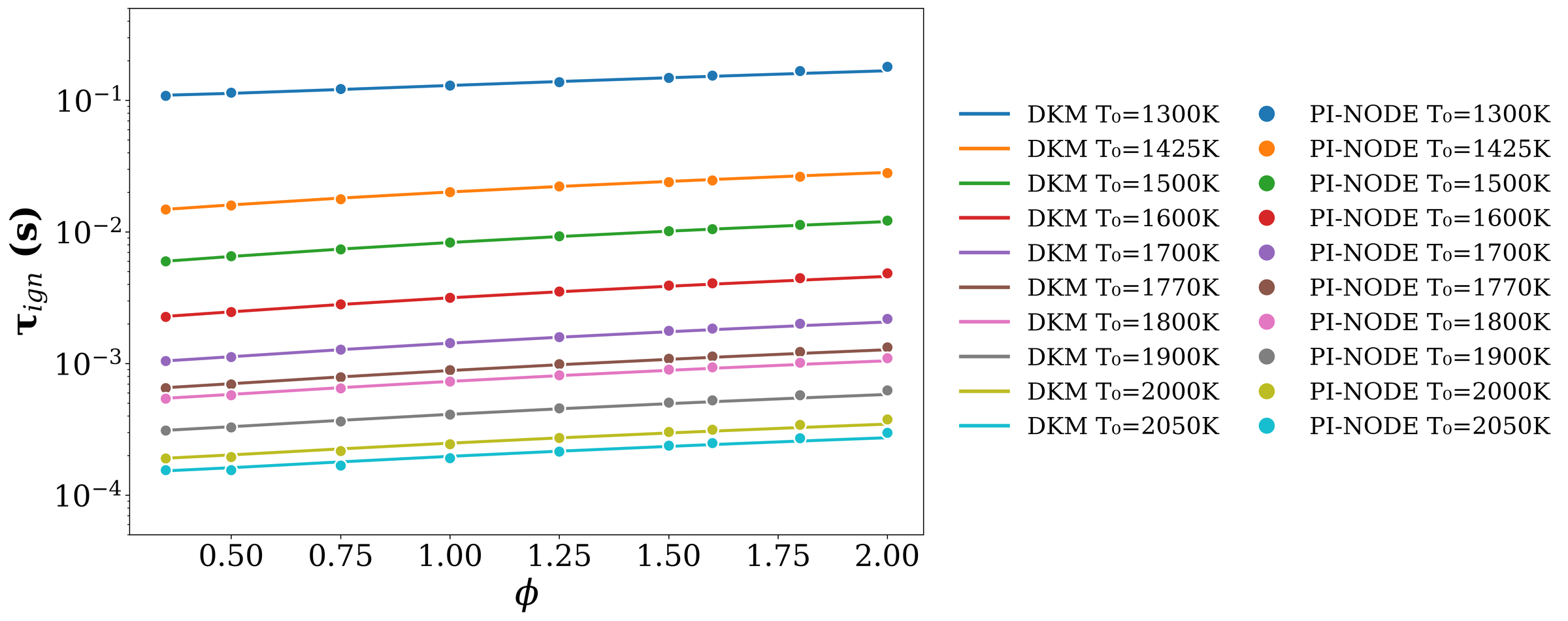}
\caption{Ignition delays of NH$_3$/air predicted by DKM (line) and PI-AE-NODE (circles) as a function of equivalence ratio for different initial temperatures.}
\label{fig:nh3_ignition_delay}
\end{figure}

Figure~\ref{fig:nh3_ignition_delay} compares the NH$_3$/air ignition delays
predicted by PI-AE-NODE and DKM over a grid that extends beyond the training
ranges to $T_0=\SI{2050}{K}$ and $\phi=2.0$. The ignition delay decreases by
nearly three orders of magnitude as $T_0$ increases from 1300 to
\SI{2050}{K} and generally increases with equivalence ratio. PI-AE-NODE
reproduces both trends, with the closest agreement within the training
domain. Larger relative differences occur at some unseen conditions, where
small deviations in the steep temperature rise shift the time at which the
ignition criterion is reached. Nevertheless, the overall mean relative error
is 1.76\%, showing that the eight-dimensional latent representation retains
the principal ignition characteristics of the 25-dimensional thermochemical
system over a wide range of timescales.

\subsubsection{Computational performance and accuracy trade-off}
\label{sec:nh3_computational_performance}

The NH$_3$/air model was benchmarked using the procedure and tolerance
selection described in Section~\ref{sec:h2_computational_performance}
(Table~\ref{tab:nh3_computational_performance}). Implicit BDF integration
provides limited benefit: PI-AE-NODE is slower than DKM in Case~1
($0.78\times$) and only marginally faster in Case~2 ($1.16\times$). At the
same tolerances, explicit RK45 integration is $2.4$--$3.5$ times faster than
BDF, giving speedups of $2.07\times$ to $2.80\times$ (Cases 3--5), and RK23
provides the highest speedup of $3.96\times$ (Case 6). The ignition delay
error remains between 1.73\% and 2.01\% in all cases, and Case~6, with an
error of 1.76\%, again offers the most favourable balance between efficiency
and accuracy. As for H$_2$/air, the advantage of explicit over implicit
integration is consistent with non-stiff latent dynamics.

The speedups for NH$_3$/air are lower than for H$_2$/air despite the larger
mechanism. This is likely related to the ten-times longer integration
interval and the larger latent dimension, which increase the cost of the
surrogate relative to that of the DKM, whose cost for this moderately sized
mechanism remains comparatively low.

\begin{table}[!htbp]
\centering
\caption{Computational performance and ignition delay accuracy of PI-AE-NODE for NH$_3$/air using different ODE solvers and tolerance settings.}
\label{tab:nh3_computational_performance}
{\fontfamily{ptm}\fontsize{9}{10.35}\selectfont
\renewcommand{\arraystretch}{1.05}
\setlength{\tabcolsep}{3pt}
\begin{tabular}{
@{}
>{\centering\arraybackslash}p{0.8cm}
>{\centering\arraybackslash}p{2.2cm}
>{\centering\arraybackslash}p{1.3cm}
>{\centering\arraybackslash}p{1.8cm}
>{\centering\arraybackslash}p{1.2cm}
>{\centering\arraybackslash}p{1.5cm}
>{\centering\arraybackslash}p{1.8cm}
@{}}
\toprule
\textbf{Case}
& \textbf{Chemistry model}
& \textbf{Solver}
& \textbf{ATOL}
& \textbf{RTOL}
& \textbf{Speedup}
& \textbf{Ignition delay error (\%)} \\
\midrule
Base & DKM        & BDF  & $10^{-12}$ ($10^{-6}$) & $10^{-8}$ & --   & --   \\
1    & PI-AE-NODE & BDF  & $10^{-5}$               & $10^{-4}$ & 0.78 & 1.75 \\
2    & PI-AE-NODE & BDF  & $10^{-4}$               & $10^{-3}$ & 1.16 & 1.78 \\
3    & PI-AE-NODE & RK45 & $10^{-6}$               & $10^{-5}$ & 2.07 & 1.75 \\
4    & PI-AE-NODE & RK45 & $10^{-5}$               & $10^{-4}$ & 2.73 & 1.73 \\
5    & PI-AE-NODE & RK45 & $10^{-4}$               & $10^{-3}$ & 2.80 & 2.01 \\
6    & PI-AE-NODE & RK23 & $10^{-4}$               & $10^{-3}$ & 3.96 & 1.76 \\
\bottomrule
\end{tabular}
}
\end{table}

\subsection{CH$_4$/air system}
\label{sec:ch4_results}

The CH$_4$/air system provides the principal scalability test for
PI-AE-NODE. After excluding Ar and its reactions, the GRI-Mech 3.0 mechanism
contains 52 species and 323 reactions, giving a 53-dimensional thermochemical
state that is mapped to ten latent variables. The resulting compression ratio
of 5.3 is considerably higher than those for H$_2$/air (2.0) and NH$_3$/air
(3.1). Methane oxidation also involves a large set of carbon-containing
radicals and intermediates with narrow concentration peaks and widely
separated chemical timescales, which must be represented within this compact
latent space.

\subsubsection{Autoencoder reconstruction performance}
\label{sec:ch4_reconstruction}

\begin{figure}[!htbp]
    \centering
    \begin{subfigure}[t]{0.51\textwidth}
        \centering
        \includegraphics[width=\linewidth]{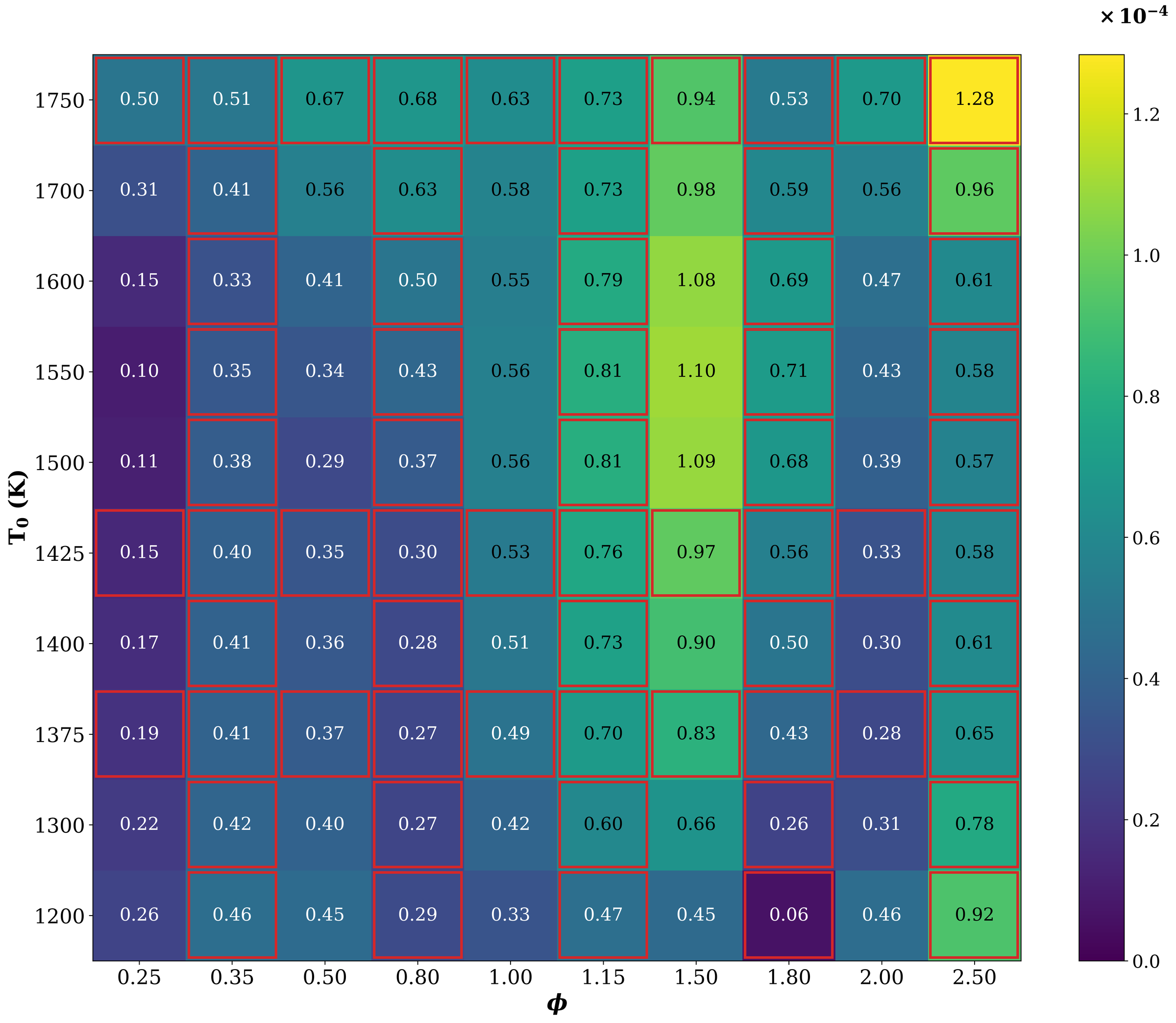}
        \caption{}
        \label{fig:ch4_species_rec}
    \end{subfigure}
    \hfill
    \begin{subfigure}[t]{0.48\textwidth}
        \centering
        \includegraphics[width=\linewidth]{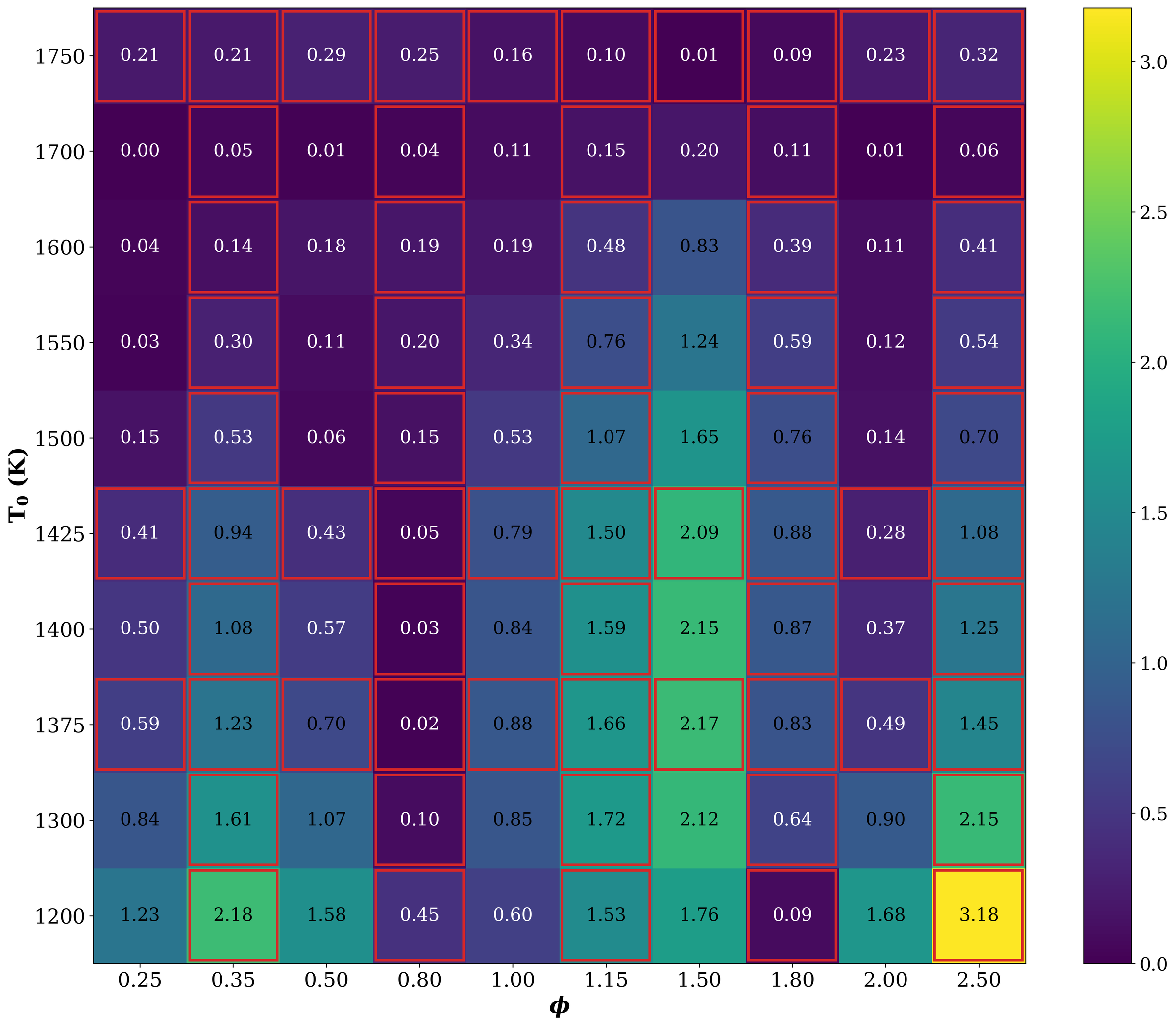}
        \caption{}
        \label{fig:ch4_temperature_rec}
    \end{subfigure}
    \caption{Reconstruction errors of the CH$_4$/air autoencoder: (a) species
    error $\varepsilon_Y$ and (b) temperature error $\varepsilon_T$ (K), as
    defined in Eq.~(\ref{eq:error_metrics}). Boxed cells denote conditions
    excluded from training.}
    \label{fig:ch4_reconstruction}
\end{figure}

The reconstruction errors in Fig.~\ref{fig:ch4_species_rec} and Fig.~\ref{fig:ch4_temperature_rec} are evaluated on a grid
that combines training conditions with unseen intermediate values and
extrapolated values. The species
reconstruction error ranges from approximately $6\times10^{-6}$ to
$1.3\times10^{-4}$, with most values of order $10^{-5}$
(Fig.~\ref{fig:ch4_species_rec}). The maximum occurs at $\phi=2.5$ and
$T_0=\SI{1750}{K}$, where both parameters lie outside the training domain, yet
it remains close to the errors at $\phi=1.5$ and $T_0=1500$--\SI{1600}{K}
within the training domain, which indicates that the rich intermediate
chemistry is the most demanding part of the composition space to compress.
The temperature reconstruction error remains below \SI{3.2}{K}
(Fig.~\ref{fig:ch4_temperature_rec}), with the largest values at the lowest initial
temperatures and a maximum of \SI{3.18}{K} at $\phi=2.5$ and
$T_0=\SI{1200}{K}$. Despite the highest compression ratio of the three
systems, the ten-dimensional latent space therefore preserves the essential
thermal and compositional features of the 53-dimensional state.

As shown in Fig.~\ref{fig:ch4_temperature_rec}, the temperature reconstruction error remains below 3.2 K, with a maximum of 3.18 K at $\phi=2.50$ and $T_0=1200$ K. Overall, the ten dimensional latent space preserves the important thermal and compositional features of the original 53 dimensional state while achieving a compression ratio of 5.3.


\subsubsection{In-domain autoignition trajectories}
\label{sec:ch4_in_domain}

\begin{figure}[!htbp]
\centering
\includegraphics[width=\textwidth]{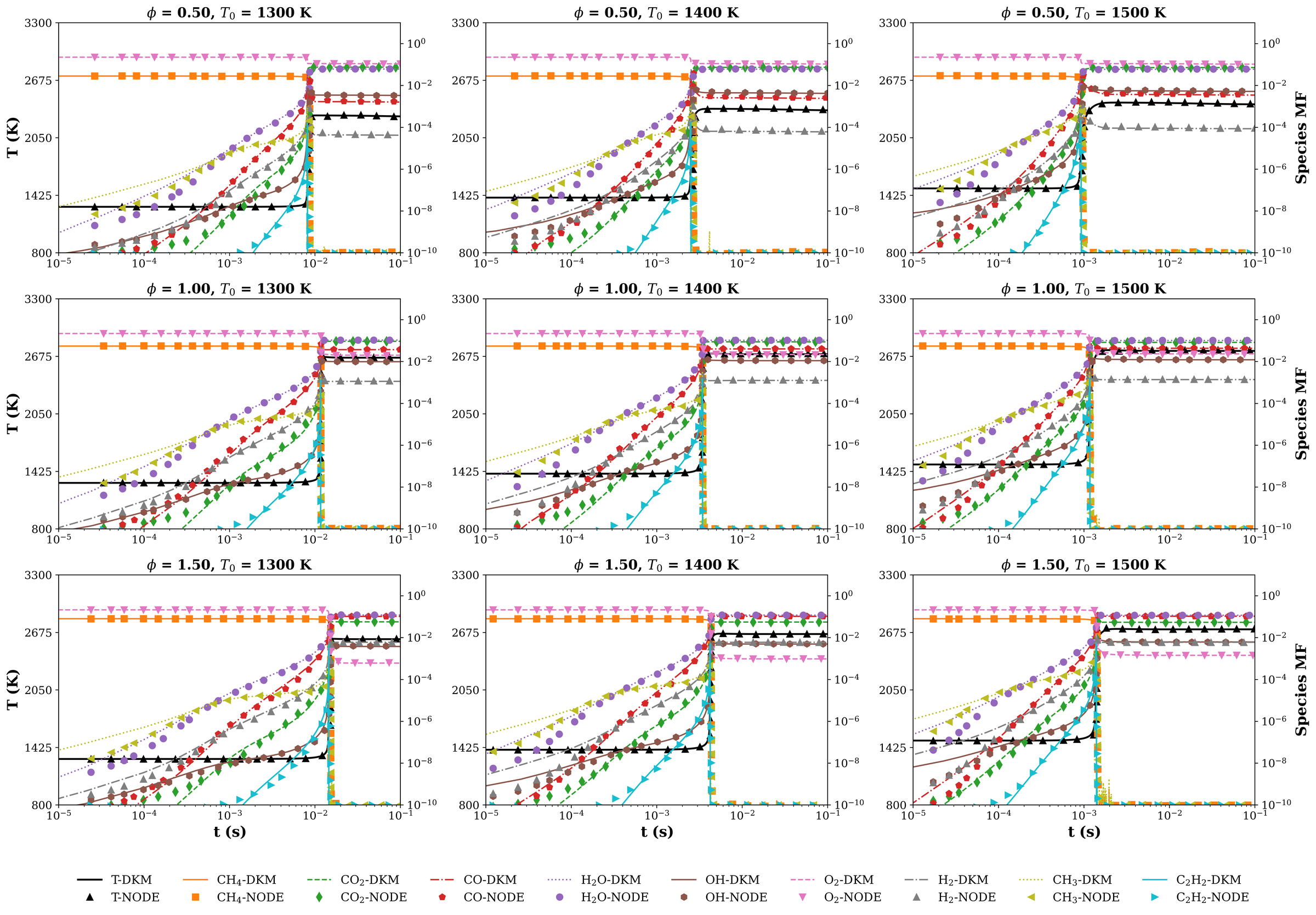}
\caption{Temperature and species mass fraction trajectories predicted by DKM
and PI-AE-NODE for CH$_4$/air at the training conditions $\phi=0.5$, 1.0, and
1.5 and $T_0=1300$, 1400, and \SI{1500}{K}.}
\label{fig:ch4_in_domain}
\end{figure}

PI-AE-NODE was evaluated at the training conditions $\phi=0.5$, 1.0, and 1.5
and $T_0=1300$, 1400, and \SI{1500}{K} (Fig.~\ref{fig:ch4_in_domain}). For
all nine conditions, the predicted temperature closely follows the DKM,
capturing the shortening of the induction period with increasing $T_0$, the
rapid temperature rise at ignition, and the dependence of the post-ignition
temperature on equivalence ratio. The predicted species profiles reproduce
CH$_4$ and O$_2$ consumption, H$_2$O and CO$_2$ formation, the production and
subsequent oxidation of CO, and the evolution of H$_2$, OH, CH$_3$, and
C$_2$H$_2$. The model therefore represents the main stages of methane
oxidation: fuel decomposition, radical formation, heat release, and oxidation
of intermediate carbon species.

\begin{figure}[!htbp]
\centering
\includegraphics[width=\textwidth,height=0.85\textheight,keepaspectratio]{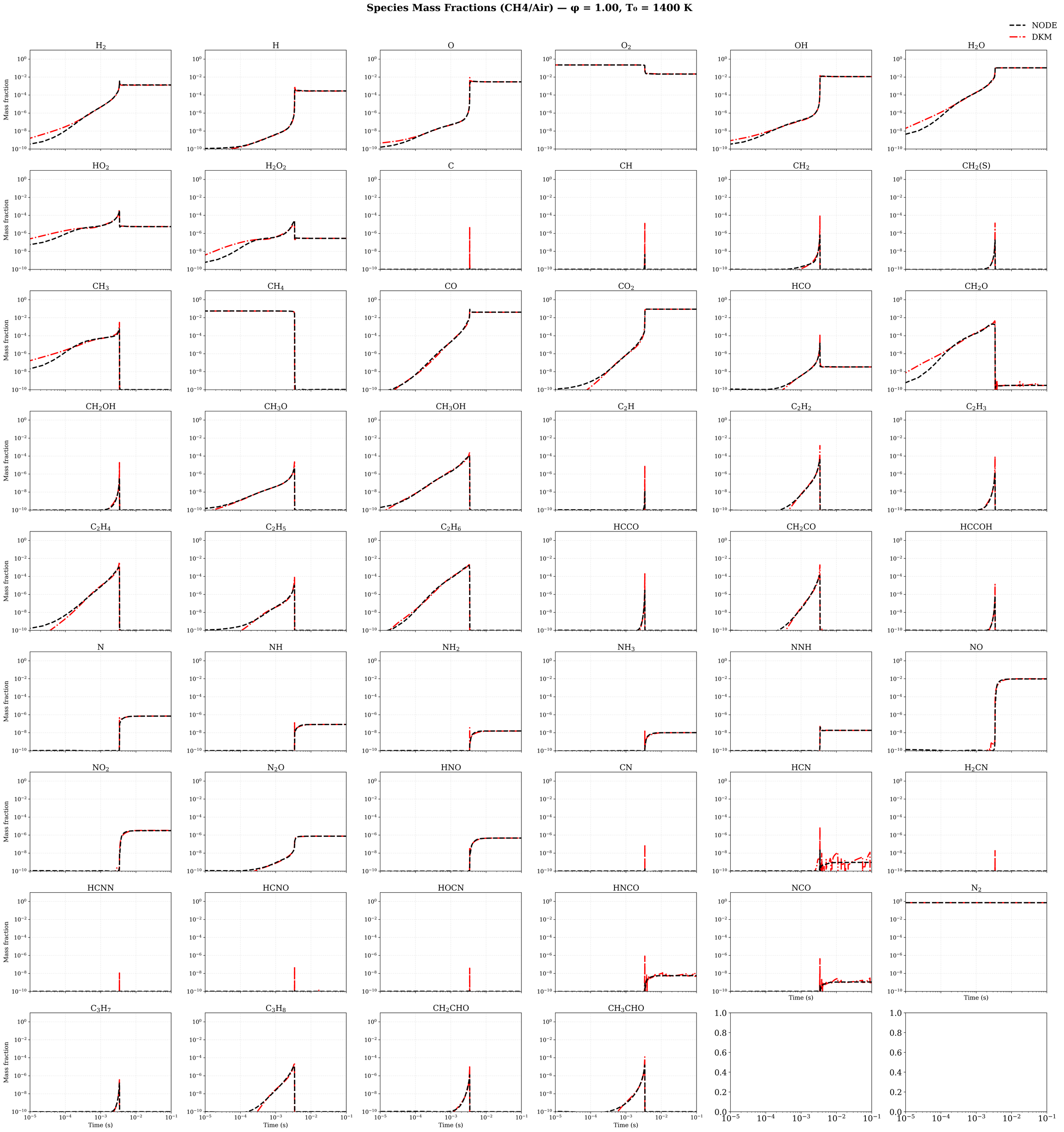}
\caption{Mass fractions of all 52 species predicted by DKM and PI-AE-NODE for
CH$_4$/air at $\phi=1.0$ and $T_0=\SI{1400}{K}$.}
\label{fig:ch4_in_domain_species}
\end{figure}

Figure~\ref{fig:ch4_in_domain_species} shows all 52 species at $\phi=1.0$
and $T_0=\SI{1400}{K}$. The major species and dominant intermediates,
including H, O, OH, HO$_2$, H$_2$O$_2$, CH$_3$, CH$_2$O, HCO, CH$_3$O, and
several C$_2$ species, agree closely with the DKM throughout the prediction
interval. Larger local differences are confined to trace species such as C,
CH, HCNN, HCNO, HOCN, and H$_2$CN, whose mass fractions remain near the floor
of the logarithmic transformation ($\varepsilon=10^{-10}$) for most of the
interval and rise only in extremely narrow peaks at ignition, which are
partially captured. These differences likely reflect the loss of small-scale
information during compression into the ten-dimensional latent space, but
they do not noticeably affect the temperature, major species, or ignition
timing.

\subsubsection{Unseen-condition autoignition trajectories}
\label{sec:ch4_unseen}

\begin{figure}[!htbp]
\centering
\includegraphics[width=\textwidth]{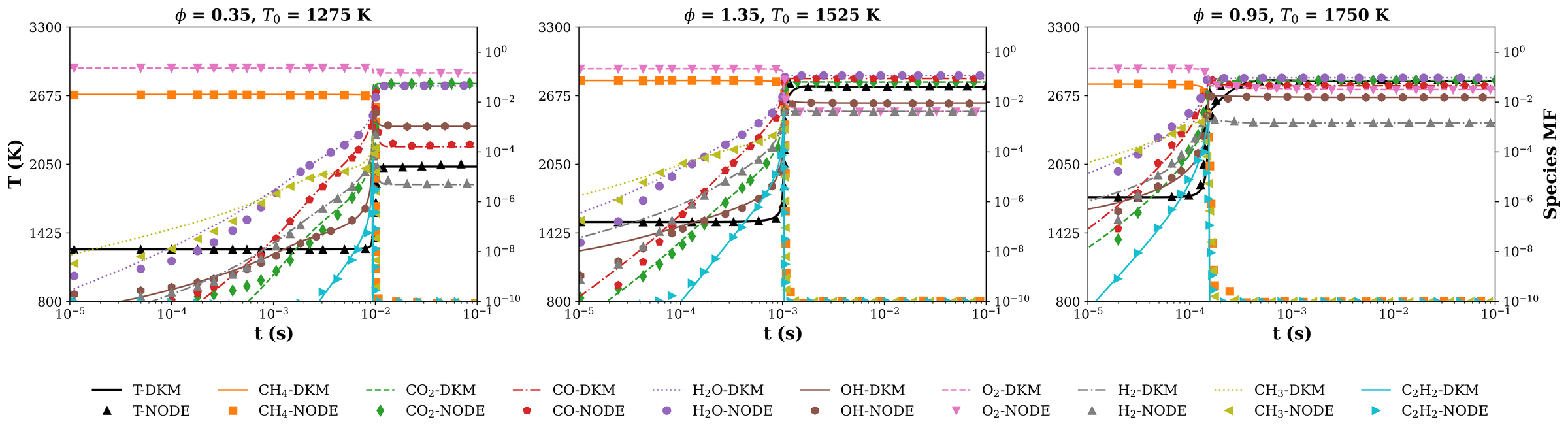}
\caption{Temperature and species mass fraction trajectories predicted by DKM
and PI-AE-NODE for CH$_4$/air at the unseen conditions
$(\phi,T_0)=(0.35,\SI{1275}{K})$ and $(1.35,\SI{1525}{K})$ (interpolated) and
$(0.95,\SI{1750}{K})$ (extrapolated).}
\label{fig:ch4_unseen}
\end{figure}

Generalization was evaluated at three conditions excluded from training
(Fig.~\ref{fig:ch4_unseen}). The conditions $(\phi,T_0)=(0.35,\SI{1275}{K})$
and $(1.35,\SI{1525}{K})$ lie between the training values, whereas
$(0.95,\SI{1750}{K})$ extends beyond the upper training limit of initial
temperature. PI-AE-NODE predicts the ignition timing and temperature
evolution at all three conditions without instability or noticeable phase
drift. At the lean condition, it captures the relatively long induction
period, CH$_4$ and O$_2$ consumption, and the formation and subsequent
oxidation of CO. At $\phi=1.35$ and $T_0=\SI{1525}{K}$, the final temperature
and major product composition are reproduced, and at the extrapolated
temperature the considerably shorter ignition timescale remains closely
aligned with the DKM. The H$_2$, OH, CH$_3$, and C$_2$H$_2$ profiles further
show that the principal radical and carbon oxidation pathways remain
represented away from the training conditions.

\begin{figure}[!htbp]
\centering
\includegraphics[width=\textwidth,height=0.85\textheight,keepaspectratio]{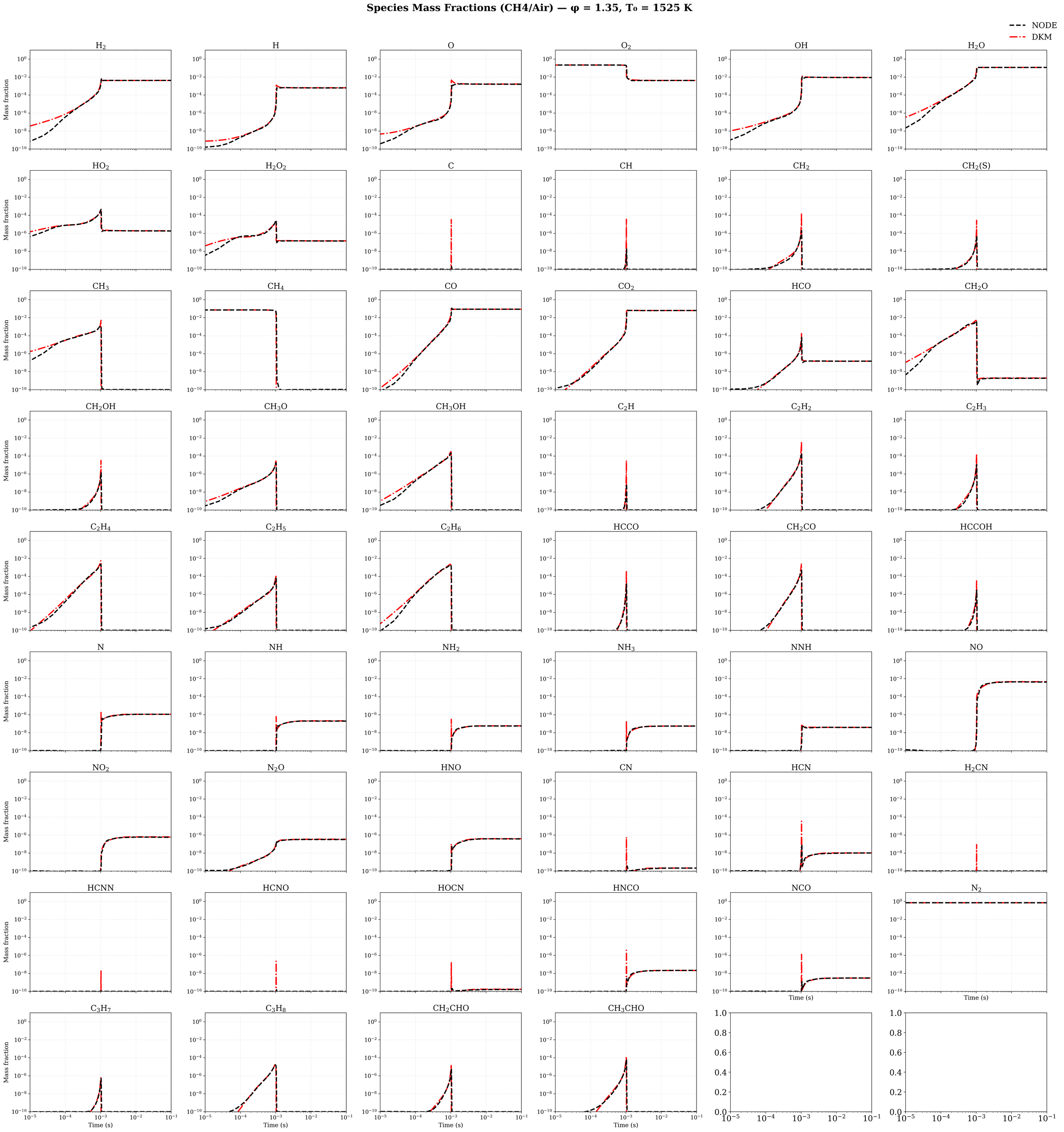}
\caption{Mass fractions of all 52 species predicted by DKM and PI-AE-NODE for
CH$_4$/air at the unseen condition $\phi=1.35$ and $T_0=\SI{1525}{K}$.}
\label{fig:ch4_unseen_species}
\end{figure}

Figure~\ref{fig:ch4_unseen_species} shows all 52 species at the interpolated
condition $\phi=1.35$ and $T_0=\SI{1525}{K}$. The major reactants, stable
products, and dominant intermediates remain closely aligned with the DKM. As
at the training condition, larger differences are confined to trace carbon
and nitrogen species that remain near the transformation floor for most of
the interval and rise only in narrow peaks at ignition. Because their mass
fractions are extremely small, these deviations have little influence on the
temperature, major species conversion, or ignition timing. The learned latent
dynamics therefore interpolate reliably between training conditions and
remain predictive under temperature extrapolation.

\subsubsection{Ignition delay prediction}
\label{sec:ch4_ignition_delay}

\begin{figure}[!htbp]
\centering
\includegraphics[width=0.7\textwidth]{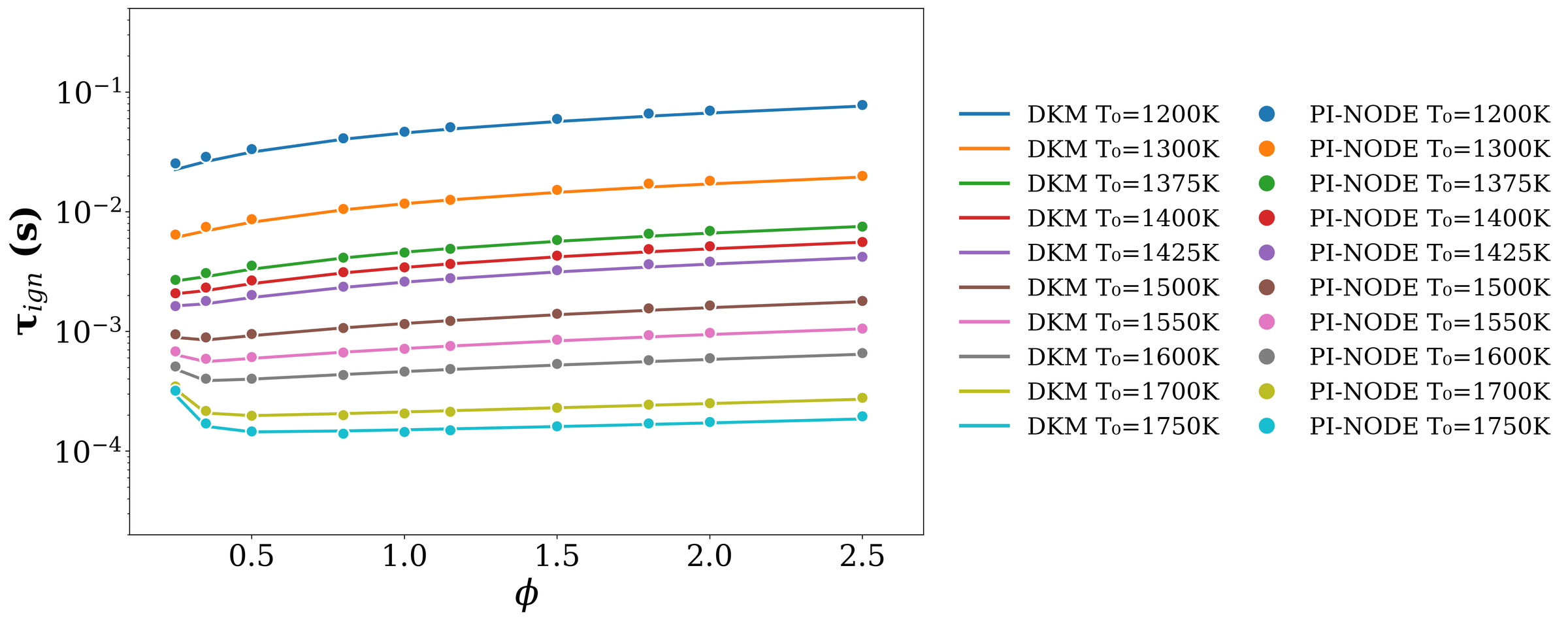}
\caption{Ignition delays of CH$_4$/air predicted by DKM (line) and PI-AE-NODE (circle) as a
function of equivalence ratio for different initial temperatures.}
\label{fig:ch4_ignition_delay}
\end{figure}

Figure~\ref{fig:ch4_ignition_delay} compares the CH$_4$/air ignition delays
predicted by PI-AE-NODE and DKM over a grid that extends beyond the training
ranges to $T_0=\SI{1750}{K}$ and $\phi=2.5$. Increasing $T_0$ from 1200 to
\SI{1750}{K} reduces the ignition delay by more than two orders of
magnitude, and the ignition delay generally increases with equivalence
ratio, with a steeper rise under very lean conditions at the highest
temperatures. PI-AE-NODE reproduces these trends, with the closest agreement
within the training domain. Larger relative differences occur at some unseen
conditions, where small deviations in the steep temperature rise shift the
time at which the ignition criterion is reached. Nevertheless, the overall
mean relative error is 3.30\%, showing that the ten-dimensional latent
representation retains the principal ignition characteristics of the
53-dimensional thermochemical system.

\subsubsection{Computational performance and accuracy trade-off}
\label{sec:ch4_computational_performance}

The CH$_4$/air model was benchmarked using the procedure of
Section~\ref{sec:h2_computational_performance}
(Table~\ref{tab:ch4_computational_performance}), with the DKM tolerances
selected by the same criterion of the loosest values that preserve accuracy
for this mechanism. Unlike for the smaller systems, PI-AE-NODE provides
substantial acceleration even with implicit BDF integration, with speedups
of $5.09\times$ and $6.66\times$ (Cases 1 and 2), which reflects the
reduction from 53 integrated variables to ten. Explicit integration provides
further acceleration: at the same tolerances, RK45 is $3.3$--$3.6$ times faster
than BDF, giving speedups of $12.74\times$ to $22.28\times$ (Cases 3--5), and
RK23 reaches $26.00\times$ (Case 6). The ignition delay error varies only
between 3.14\% and 3.31\% across all cases, so Case~6, with an error of
3.30\%, provides the most favorable trade-off. As for the smaller systems,
the advantage of explicit integration is consistent with non-stiff latent
dynamics.

\begin{table}[!htbp]
\centering
\caption{Computational performance and ignition delay accuracy of PI-AE-NODE for CH$_4$/air using different ODE solvers and tolerance settings.}
\label{tab:ch4_computational_performance}
{\fontfamily{ptm}\fontsize{9}{10.35}\selectfont
\renewcommand{\arraystretch}{1.05}
\setlength{\tabcolsep}{3pt}
\begin{tabular}{
@{}
>{\centering\arraybackslash}p{0.8cm}
>{\centering\arraybackslash}p{2.2cm}
>{\centering\arraybackslash}p{1.3cm}
>{\centering\arraybackslash}p{1.8cm}
>{\centering\arraybackslash}p{1.2cm}
>{\centering\arraybackslash}p{1.5cm}
>{\centering\arraybackslash}p{1.8cm}
@{}}
\toprule
\textbf{Case}
& \textbf{Chemistry model}
& \textbf{Solver}
& \textbf{ATOL}
& \textbf{RTOL}
& \textbf{Speedup}
& \textbf{Ignition delay error (\%)} \\
\midrule
Base & DKM        & BDF  & $10^{-10}$ ($10^{-5}$) & $10^{-6}$ & --    & --   \\
1    & PI-AE-NODE & BDF  & $10^{-5}$               & $10^{-4}$ & 5.09  & 3.26 \\
2    & PI-AE-NODE & BDF  & $10^{-4}$               & $10^{-3}$ & 6.66  & 3.14 \\
3    & PI-AE-NODE & RK45 & $10^{-6}$               & $10^{-5}$ & 12.74 & 3.28 \\
4    & PI-AE-NODE & RK45 & $10^{-5}$               & $10^{-4}$ & 18.56 & 3.27 \\
5    & PI-AE-NODE & RK45 & $10^{-4}$               & $10^{-3}$ & 22.28 & 3.31 \\
6    & PI-AE-NODE & RK23 & $10^{-4}$               & $10^{-3}$ & 26.00 & 3.30 \\
\bottomrule
\end{tabular}
}
\end{table}

\subsubsection{Computational performance and accuracy trade-off}
\label{sec:ch4_computational_performance}

The CH$_4$/air model was benchmarked using the procedure of
Section~\ref{sec:h2_computational_performance}
(Table~\ref{tab:ch4_computational_performance}), with the DKM tolerances
selected by the same criterion of the loosest values that preserve accuracy
for this mechanism. Unlike for the smaller systems, PI-AE-NODE provides
substantial acceleration even with implicit BDF integration, with speedups
of $5.09\times$ and $6.66\times$ (Cases 1 and 2), which reflects the
reduction from 53 integrated variables to ten. Explicit integration provides
further acceleration: at the same tolerances, RK45 is $3.3$--$3.6$ times faster
than BDF, giving speedups of $12.74\times$ to $22.28\times$ (Cases 3--5), and
RK23 reaches $26.00\times$ (Case 6). The ignition delay error varies only
between 3.14\% and 3.31\% across all cases, so Case~6, with an error of
3.30\%, provides the most favourable trade-off. As for the smaller systems,
the advantage of explicit integration is consistent with non-stiff latent
dynamics.

\subsection{Comparison across fuels and limitations}
\label{sec:cross_fuel}

Across the three fuel systems, the latent dimension increases from 5 to 10
while the physical state dimension increases from 10 to 53, and the mean
ignition delay errors remain between 1.76\% and 3.30\%. The maximum speedup,
$5.69\times$, $3.96\times$, and $26.00\times$ for H$_2$/air, NH$_3$/air, and
CH$_4$/air, respectively, does not increase uniformly with mechanism size,
because it depends on both the cost of the DKM and that of the surrogate. The
cost of implicit DKM integration grows rapidly with the number of species,
through the evaluation of the reaction rates and the Jacobian-based linear
algebra, whereas the cost of PI-AE-NODE is dominated by its hidden layers of
fixed width and depends only weakly on the latent dimension. This favours
the surrogate most strongly for the largest mechanism. For the moderately
sized NH$_3$/air mechanism, the longer integration interval and larger latent
dimension relative to H$_2$/air reduce the net benefit, so the advantage of
PI-AE-NODE becomes decisive only for larger mechanisms.

The present framework has been trained and evaluated using homogeneous,
adiabatic, constant-pressure reactor trajectories at atmospheric pressure. Its
performance under variable pressure, heat loss, dilution, and direct coupling
with multidimensional reacting-flow solvers has not yet been assessed.
Although the major species and ignition behaviour are predicted accurately,
larger local deviations occur for trace species with extremely low mass
fractions and narrow concentration peaks, particularly under extrapolation.
These deviations do not noticeably affect the temperature, major species
conversion, or ignition delay within the conditions examined, but they define
the main directions for improvement discussed in
Section~\ref{sec:conclusions}.

\section{Conclusions and future work}
\label{sec:conclusions}

A physics-informed autoencoder Neural ODE (PI-AE-NODE) framework was
developed as a reduced-order surrogate for detailed combustion kinetics. The
framework combines a logarithmic transformation and standardization of the
thermochemical state, an autoencoder that compresses all species and
temperature into a low-dimensional latent space, and a Neural ODE that
advances the latent state continuously in time. Elemental conservation, mass
conservation, and enthalpy consistency were imposed on both the
time-integrated trajectories and the direct reconstruction. The framework was
evaluated for H$_2$/air, NH$_3$/air, and CH$_4$/air mechanisms containing 9,
24, and 52 species and 21, 100, and 323 reactions, whose 10-, 25-, and
53-dimensional states were represented by 5, 8, and 10 latent variables,
respectively.

The main findings are as follows. For H$_2$/air, the physical constraints
reduced the reconstruction errors, the conservation errors near ignition, and
the mean ignition delay error, which decreased from 6.19\% for an otherwise
identical data-driven model to 2.20\%. The autoencoders retained the principal
compositional and thermal features of all three systems, with maximum
temperature reconstruction errors of approximately 11.1, 4.06, and
\SI{3.18}{K} for H$_2$/air, NH$_3$/air, and CH$_4$/air, respectively.
PI-AE-NODE reproduced the induction period, heat release, major and
intermediate species evolution, and post-ignition state across lean,
stoichiometric, and rich mixtures, including at unseen conditions, and
retained the dependence of ignition delay on initial temperature and
equivalence ratio. Larger local deviations were confined to trace species and
to strong extrapolation beyond the training domain. Explicit integration of
the learned latent dynamics was consistently more efficient than implicit BDF
integration, consistent with non-stiff latent dynamics. With RK23, the
maximum speedups were $5.69\times$, $3.96\times$, and $26.00\times$ for
H$_2$/air, NH$_3$/air, and CH$_4$/air, with corresponding mean ignition delay
errors of 2.20\%, 1.76\%, and 3.30\%. The latent dimension increased only
modestly relative to the physical state dimension, and the computational
advantage was most pronounced for the largest mechanism.

Future work should extend the training domain to improve predictions under
strong extrapolation and improve the representation of trace species with
narrow concentration peaks, for example through adaptive sampling near
ignition and in regions of high prediction uncertainty and through improved
weighting of minor species. The framework should also be assessed under
variable pressure, heat loss, and dilution, and for broader fuel
compositions, larger kinetic mechanisms, and pollutant formation pathways.
Benchmarking on GPUs will be used to evaluate the potential for parallel
acceleration. Finally, coupling PI-AE-NODE with multidimensional
reacting-flow solvers through operator splitting is required to assess its
accuracy, stability, and computational benefit under practical CFD
conditions.

\section*{CRediT authorship contribution statement}

\noindent
Jay Patel: Conceptualization, Methodology, Software, Validation,
Formal analysis, Investigation, Data curation, Visualization,
Writing -- original draft.

\noindent
Shubhangi Bansude: Conceptualization, Methodology, Supervision,
Project administration, Resources, Writing -- review and editing.

\section*{Declaration of competing interest}

The authors declare that they have no known competing financial
interests or personal relationships that could have appeared to
influence the work reported in this paper.

\bibliographystyle{unsrtnat}
\bibliography{references}

@misc{cantera,
    author = {David G. Goodwin and Raymond L. Speth and Harry K. Moffat
              and Bryan W. Weber},
    title = {Cantera: An Object-oriented Software Toolkit for Chemical
             Kinetics, Thermodynamics, and Transport Processes},
    year = 2021,
    note = "Version 2.5.1",
    howpublished = "\url{https://www.cantera.org}",
    doi = {10.5281/zenodo.4527812}
}

@article{Sun2025AmmoniaReview,
  author  = {Sun, Jihao and Zhao, Ningbo and Zheng, Hongtao},
  title   = {A comprehensive review of ammonia combustion: Fundamental characteristics, chemical kinetics, and applications in energy systems},
  journal = {Fuel},
  volume  = {394},
  pages   = {135135},
  year    = {2025},
  doi     = {10.1016/j.fuel.2025.135135}
}

@article{Lu2009RealisticFuelChemistry,
  author  = {Lu, Tianfeng and Law, Chung K.},
  title   = {Toward accommodating realistic fuel chemistry in large-scale computations},
  journal = {Progress in Energy and Combustion Science},
  volume  = {35},
  number  = {2},
  pages   = {192--215},
  year    = {2009},
  doi     = {10.1016/j.pecs.2008.10.002}
}

@misc{Bansude2026PhysicsGuidedML,
  author        = {Bansude, Shubhangi and Patel, Jay},
  title         = {Physics-Guided Machine Learning for Predictive Turbulent Combustion Simulation},
  year          = {2026},
  eprint        = {2608.29033},
  archiveprefix = {arXiv},
  primaryclass  = {physics.flu-dyn},
  url           = {https://arxiv.org/abs/2608.29033}
}

@article{Lu2005DRG,
  author  = {Lu, Tianfeng and Law, Chung K.},
  title   = {A directed relation graph method for mechanism reduction},
  journal = {Proceedings of the Combustion Institute},
  volume  = {30},
  number  = {1},
  pages   = {1333--1341},
  year    = {2005},
  doi     = {10.1016/j.proci.2004.08.145}
}

@article{Lu2006QSSA,
  author  = {Lu, Tianfeng and Law, Chung K.},
  title   = {Systematic approach to obtain analytic solutions of quasi steady state species in reduced mechanisms},
  journal = {The Journal of Physical Chemistry A},
  volume  = {110},
  number  = {49},
  pages   = {13202--13208},
  year    = {2006},
  doi     = {10.1021/jp064482y}
}

@article{Valorani2006CSP,
  author  = {Valorani, Mauro and Creta, Francesco and Goussis, Dimitris A. and Lee, Jeremiah C. and Najm, Habib N.},
  title   = {An automatic procedure for the simplification of chemical kinetic mechanisms based on {CSP}},
  journal = {Combustion and Flame},
  volume  = {146},
  number  = {1--2},
  pages   = {29--51},
  year    = {2006},
  doi     = {10.1016/j.combustflame.2006.03.011}
}

@article{Maas1992ILDM,
  author  = {Maas, U. and Pope, S. B.},
  title   = {Simplifying chemical kinetics: Intrinsic low-dimensional manifolds in composition space},
  journal = {Combustion and Flame},
  volume  = {88},
  number  = {3--4},
  pages   = {239--264},
  year    = {1992},
  doi     = {10.1016/0010-2180(92)90034-M}
}

@article{Pope1997ISAT,
  author  = {Pope, S. B.},
  title   = {Computationally efficient implementation of combustion chemistry using in situ adaptive tabulation},
  journal = {Combustion Theory and Modelling},
  volume  = {1},
  number  = {1},
  pages   = {41--63},
  year    = {1997},
  doi     = {10.1080/713665229}
}

@inproceedings{Chen2000ANNISAT,
  author    = {Chen, J.-Y. and Blasco, J. A. and Fueyo, N. and Dopazo, C.},
  title     = {An economical strategy for storage of chemical kinetics: Fitting in situ adaptive tabulation with artificial neural networks},
  booktitle = {Proceedings of the Combustion Institute},
  volume    = {28},
  pages     = {115--121},
  year      = {2000}
}

@article{Cai2026GPUSolver,
  author  = {Cai, Yuqing and Ban, Yangyang and Dang, Guanlin and Gong, Tao and Wang, Hu and Zhang, Fan and Chen, Zhi X.},
  title   = {A {GPU}-accelerated stiff {ODE} solver for combustion {CFD} with fully analytical chemical kinetics and heterogeneous {CPU}/{GPU} integration},
  journal = {Fuel},
  volume  = {408},
  pages   = {137627},
  year    = {2026},
  doi     = {10.1016/j.fuel.2025.137627}
}

@article{Ihme2022CombustionML,
  author  = {Ihme, Matthias and Chung, Wai Tong and Mishra, Aashwin Ananda},
  title   = {Combustion machine learning: Principles, progress and prospects},
  journal = {Progress in Energy and Combustion Science},
  volume  = {91},
  pages   = {101010},
  year    = {2022},
  doi     = {10.1016/j.pecs.2022.101010}
}

@article{Ranade2019HybridChemistry,
  author  = {Ranade, Rishikesh and Alqahtani, Sultan and Farooq, Aamir and Echekki, Tarek},
  title   = {An {ANN} based hybrid chemistry framework for complex fuels},
  journal = {Fuel},
  volume  = {241},
  pages   = {625--636},
  year    = {2019},
  doi     = {10.1016/j.fuel.2018.12.082}
}

@article{Chi2021OnTheFlyANN,
  author  = {Chi, Cheng and Janiga, G{\'a}bor and Th{\'e}venin, Dominique},
  title   = {On-the-fly artificial neural network for chemical kinetics in direct numerical simulations of premixed combustion},
  journal = {Combustion and Flame},
  volume  = {226},
  pages   = {467--477},
  year    = {2021},
  doi     = {10.1016/j.combustflame.2020.12.038}
}

@article{Zhang2022MultiscaleSampling,
  author  = {Zhang, Tianhan and Yi, Yuxiao and Xu, Yifan and Chen, Zhi X. and Zhang, Yaoyu and E, Weinan and Xu, Zhi Qin John},
  title   = {A multi-scale sampling method for accurate and robust deep neural network to predict combustion chemical kinetics},
  journal = {Combustion and Flame},
  volume  = {245},
  pages   = {112319},
  year    = {2022},
  doi     = {10.1016/j.combustflame.2022.112319}
}

@article{Kumar2024DeepONets,
  author  = {Kumar, Anuj and Echekki, Tarek},
  title   = {Combustion chemistry acceleration with {DeepONets}},
  journal = {Fuel},
  volume  = {365},
  pages   = {131212},
  year    = {2024},
  doi     = {10.1016/j.fuel.2024.131212}
}

@article{Bai2024MetaLearning,
  author  = {Bai, Tianzi and Huai, Ying and Liu, Tingting and Jia, Shuqin and You, Meng and Chang, Naiwen},
  title   = {Acceleration of the complex reacting flow simulation with a generalizable neural network based on meta-learning},
  journal = {Fuel},
  volume  = {372},
  pages   = {132173},
  year    = {2024},
  doi     = {10.1016/j.fuel.2024.132173}
}

@inproceedings{Chen2018NeuralODE,
  author    = {Chen, Ricky T. Q. and Rubanova, Yulia and Bettencourt, Jesse and Duvenaud, David},
  title     = {Neural Ordinary Differential Equations},
  booktitle = {Advances in Neural Information Processing Systems},
  volume    = {31},
  year      = {2018},
  url       = {https://arxiv.org/abs/1806.07366}
}

@article{Owoyele2022ChemNODE,
  author  = {Owoyele, Opeoluwa and Pal, Pinaki},
  title   = {{ChemNODE}: A neural ordinary differential equations framework for efficient chemical kinetic solvers},
  journal = {Energy and AI},
  volume  = {7},
  pages   = {100118},
  year    = {2022},
  doi     = {10.1016/j.egyai.2021.100118}
}

@article{Bansude2023DataDrivenNODE,
  author  = {Bansude, Shubhangi and Imani, Farhad and Sheikhi, Reza},
  title   = {A Data-Driven Framework for Computationally Efficient Integration of Chemical Kinetics Using Neural Ordinary Differential Equations},
  journal = {ASME Open Journal of Engineering},
  volume  = {2},
  pages   = {021022},
  year    = {2023},
  doi     = {10.1115/1.4062105}
}

@article{Bansude2023PMSR,
  author  = {Bansude, Shubhangi and Imani, Farhad and Sheikhi, Reza},
  title   = {Performance Assessment of Chemical Kinetics Neural Ordinary Differential Equations in Pairwise Mixing Stirred Reactor},
  journal = {ASME Open Journal of Engineering},
  volume  = {2},
  pages   = {021003},
  year    = {2023},
  doi     = {10.1115/1.4056476}
}

@article{Saito2023AmmoniaNODE,
  author  = {Saito, Manabu and Xing, Jiangkuan and Nagao, Jun and Kurose, Ryoichi},
  title   = {Data-driven simulation of ammonia combustion using neural ordinary differential equations ({NODE})},
  journal = {Applications in Energy and Combustion Science},
  volume  = {16},
  pages   = {100196},
  year    = {2023},
  doi     = {10.1016/j.jaecs.2023.100196}
}

@inproceedings{Dikeman2022StiffnessReducedNODE,
  author    = {Dikeman, Henry E. and Zhang, Hongyuan and Yang, Suo},
  title     = {Stiffness-Reduced Neural {ODE} Models for Data-Driven Reduced-Order Modeling of Combustion Chemical Kinetics},
  booktitle = {AIAA SCITECH 2022 Forum},
  publisher = {American Institute of Aeronautics and Astronautics},
  year      = {2022},
  doi       = {10.2514/6.2022-0226}
}

@article{Vijayarangan2024DynamicsInformed,
  author  = {Vijayarangan, Vijayamanikandan and Uranakara, Harshavardhana A. and Barwey, Shivam and Galassi, Riccardo Malpica and Malik, Mohammad Rafi and Valorani, Mauro and Raman, Venkat and Im, Hong G.},
  title   = {A data-driven reduced-order model for stiff chemical kinetics using dynamics-informed training},
  journal = {Energy and AI},
  volume  = {15},
  pages   = {100325},
  year    = {2024},
  doi     = {10.1016/j.egyai.2023.100325}
}

@article{Zhang2024CRKPINN,
  author  = {Zhang, Shihong and Zhang, Chi and Wang, Bosen},
  title   = {{CRK-PINN}: A physics-informed neural network for solving combustion reaction kinetics ordinary differential equations},
  journal = {Combustion and Flame},
  volume  = {269},
  pages   = {113647},
  year    = {2024},
  doi     = {10.1016/j.combustflame.2024.113647}
}

@article{Wang2025PhysicalConservation,
  author  = {Wang, Tinghao and Yi, Yuxiao and Yao, Junjie and Xu, Zhi Qin John and Zhang, Tianhan and Chen, Zheng},
  title   = {Enforcing physical conservation in neural network surrogate models for complex chemical kinetics},
  journal = {Combustion and Flame},
  volume  = {275},
  pages   = {114105},
  year    = {2025},
  doi     = {10.1016/j.combustflame.2025.114105}
}

@article{Kumar2025PCNODE,
  author  = {Kumar, Tadbhagya and Kumar, Anuj and Pal, Pinaki},
  title   = {A physics-constrained neural ordinary differential equations approach for robust learning of stiff chemical kinetics},
  journal = {Combustion Theory and Modelling},
  volume  = {29},
  number  = {3},
  pages   = {332--347},
  year    = {2025},
  doi     = {10.1080/13647830.2025.2478266}
}

@article{Kumar2025PhyChemNODE,
  author  = {Kumar, Tadbhagya and Kumar, Anuj and Pal, Pinaki},
  title   = {{Phy-ChemNODE}: An end-to-end physics-constrained autoencoder-{NeuralODE} framework for learning stiff chemical kinetics of hydrocarbon fuels},
  journal = {Frontiers in Thermal Engineering},
  volume  = {5},
  pages   = {1594443},
  year    = {2025},
  doi     = {10.3389/fther.2025.1594443}
}

@article{Boivin2011HydrogenMechanism,
  author  = {Boivin, P. and Jim{\'e}nez, C. and S{\'a}nchez, A. L. and Williams, F. A.},
  title   = {An explicit reduced mechanism for {H2}-air combustion},
  journal = {Proceedings of the Combustion Institute},
  volume  = {33},
  number  = {1},
  pages   = {517--523},
  year    = {2011},
  doi     = {10.1016/j.proci.2010.05.002}
}

@misc{UCSDChemicalMechanisms,
  author       = {{University of California San Diego}},
  title        = {Chemical-Kinetic Mechanisms for Combustion Applications},
  year         = {n.d.},
  howpublished = {Combustion Research Group, University of California San Diego},
  url          = {https://web.eng.ucsd.edu/mae/groups/combustion/mechanism.html},
  note         = {Accessed: 13 October 2026}
}

@article{Okafor2018MethaneAmmonia,
  author  = {Okafor, Ekenechukwu C. and Naito, Yuji and Colson, Sophie and Ichikawa, Akinori and Kudo, Taku and Hayakawa, Akihiro and Kobayashi, Hideaki},
  title   = {Experimental and numerical study of the laminar burning velocity of {CH4--NH3--air} premixed flames},
  journal = {Combustion and Flame},
  volume  = {187},
  pages   = {185--198},
  year    = {2018},
  doi     = {10.1016/j.combustflame.2017.09.002}
}

@misc{Smith1999GRIMech30,
  author       = {Smith, Gregory P. and Golden, David M. and Frenklach, Michael and Moriarty, Nigel W. and Eiteneer, Boris and Goldenberg, Mikhail and Bowman, C. Thomas and Hanson, Ronald K. and Song, Soonho and Gardiner, William C. Jr. and Lissianski, Vitali V. and Qin, Zhiwei},
  title        = {{GRI-Mech} 3.0},
  year         = {1999},
  howpublished = {Gas Research Institute},
  url          = {https://combustion.berkeley.edu/gri-mech/version30/text30.html},
  note         = {Accessed: 13 October 2026}
}

\end{document}